\documentstyle[12pt,epsf]{article}
\newcommand{\be}{\begin{equation}}
\newcommand{\ee}{\end{equation}}

\newcommand{\beqa}{\begin{eqnarray}}
\newcommand{\eeqa}{\end{eqnarray}}
\newcommand{\nn}{\nonumber}

\newcommand{\eqref}[1]{(\ref{#1})}

\def\boxit#1{\vbox{\hrule\hbox{\vrule\kern8pt
\vbox{\hbox{\kern8pt}\hbox{\vbox{#1}}\hbox{\kern8pt}}
\kern8pt\vrule}\hrule}}
\def\mathboxit#1{\vbox{\hrule\hbox{\vrule\kern8pt\vbox{\kern8pt
\hbox{$\displaystyle #1$}\kern8pt}\kern8pt\vrule}\hrule}}

\def\IB{\relax\hbox{$\inbar\kern-.3em{\rm B}$}}
\def\IC{\relax\hbox{$\inbar\kern-.3em{\rm C}$}}
\def\ID{\relax\hbox{$\inbar\kern-.3em{\rm D}$}}
\def\IE{\relax\hbox{$\inbar\kern-.3em{\rm E}$}}
\def\IF{\relax\hbox{$\inbar\kern-.3em{\rm F}$}}
\def\IG{\relax\hbox{$\inbar\kern-.3em{\rm G}$}}
\def\IGa{\relax\hbox{${\rm I}\kern-.18em\Gamma$}}
\def\IH{\relax{\rm I\kern-.18em H}}
\def\IK{\relax{\rm I\kern-.18em K}}
\def\IL{\relax{\rm I\kern-.18em L}}
\def\IP{\relax{\rm I\kern-.18em P}}
\def\IR{\relax{\rm I\kern-.18em R}}
\def\IZ{\relax\ifmmode\mathchoice
{\hbox{\cmss Z\kern-.4em Z}}{\hbox{\cmss Z\kern-.4em Z}}
{\lower.9pt\hbox{\cmsss Z\kern-.4em Z}} {\lower1.2pt\hbox{\cmsss
Z\kern-.4em Z}}\else{\cmss Z\kern-.4em Z}\fi}

\def\II{\relax{\rm I\kern-.18em I}}

\def\CA {{\cal A}}

\def\CG {{\cal G}}

\def\CL {{\cal L}}

\begin{document}

\hfill  NRCPS-HE-06-35

\vspace{1cm}
\begin{center}
{\LARGE ~\\ Non-Abelian Tensor Gauge Fields}\\

{\large ~\\Enhanced Symmetries

~\\

}

\vspace{1cm}

{\sl George Savvidy\\
Demokritos National Research Center\\
Institute of Nuclear Physics\\
Plato Laboratory for Theoretical Physics\\
Ag. Paraskevi, GR-15310 Athens,Greece  \\
\centerline{\footnotesize\it E-mail: savvidy@inp.demokritos.gr}
}
\end{center}
\vspace{60pt}

\centerline{{\bf Abstract}}

\vspace{12pt}

\noindent
We define a group of extended non-Abelian gauge transformations for
tensor gauge fields.
On this group one can define generalized field strength tensors,
which are transforming homogeneously with respect to the extended
gauge transformations.
The generalized field strength tensors allow to
construct two infinite series of gauge invariant quadratic forms.
Each term of these infinite series is separately gauge invariant.
The invariant Lagrangian is a linear sum of these forms
and describes interaction of tensor gauge
fields of arbitrarily large integer spins $1,2,...$. It does not contain
higher derivatives of the tensor gauge fields, and all interactions take place
through three- and four-particle exchanges with dimensionless coupling constant.
The first term in this sum is the Yang-Mills Lagrangian.

The invariance with respect to
the extended gauge transformations does not fix
the coefficients - the coupling constants - in front of these forms.
There is a freedom to vary them without breaking the extended gauge symmetry.
We demonstrate that by an appropriate tuning of these coupling constants
one can achieve an enhancement of the extended gauge symmetry.
This leads to highly symmetric equations. We present
the explicit form of the free equations
for the rank-2  and rank-3 gauge fields.
Their relation to the Schwinger free equation for the rank-3 gauge fields is discussed.


\newpage

\pagestyle{plain}

\tableofcontents

\section{\it Introduction}

It is well understood, that the concept of local gauge
invariance allows to define non-Abelian gauge fields \cite{yang},
to derive their  dynamical field equations
and to develop a universal point of view on matter interactions
as resulting from the exchange of gauge quanta of different forms.
It is appealing to extend the gauge principle so that it will define the
interaction of matter fields which carry not only non-commutative internal charges, but
also arbitrary half-integer spins. This extension
will induce the interaction of matter fields mediated by a charged
gauge quanta carrying a spin larger than one \cite{Savvidy:2005zm,Savvidy:2005fi}.

In our recent approach  the gauge fields are defined as
rank-$(s+1)$ tensors
\cite{Savvidy:2005zm,Savvidy:2005fi,Savvidy:2005ki,Savvidy:2005vm,Savvidy:2005at}
$$
A^{a}_{\mu\lambda_1 ... \lambda_{s}}
$$
and are totally symmetric with respect to the
indices $  \lambda_1 ... \lambda_{s}  $.  A priory the tensor fields
have no symmetries with
respect to the first index  $\mu$. This is an essential departure from the
previous considerations, in which the higher-rank tensors were totally symmetric
\cite{fierz,fierzpauli,minkowski,yukawa1,wigner,schwinger,Weinberg:1964cn,chang,singh,fronsdal}.
The index $s$ runs from zero to infinity.
The first member of this family of the tensor gauge bosons is the Yang-Mills
vector boson $A^{a}_{\mu}$.

The extended non-Abelian gauge transformation of the tensor gauge fields
\cite{Savvidy:2005zm,Savvidy:2005fi}
$$
\delta_{\xi} ~A^{a}_{\mu\lambda_1\lambda_2 ...\lambda_s}
$$
is defined
by the equation (\ref{polygauge}) and
comprise a closed algebraic structure, because
the commutator of two transformations can be expressed in the form
$$
[~\delta_{\eta},\delta_{\xi}]~A_{\mu\lambda_1\lambda_2 ...\lambda_s} ~=~
-i g~ \delta_{\zeta} A_{\mu\lambda_1\lambda_2 ...\lambda_s}
$$
where the gauge parameters $\{\zeta\}$ are given by the matrix commutators (\ref{gaugealgebra}).
This allows to define generalized field strength tensors (\ref{fieldstrengthparticular})
\cite{Savvidy:2005zm,Savvidy:2005fi}
$$
G^{a}_{\mu\nu,\lambda_{1}...\lambda_{s}}
$$
which are {\it transforming homogeneously} (\ref{fieldstrenghparticulartransformation})
with respect to the extended gauge transformations (\ref{polygauge}).

The field strength tensors $G^{a}_{\mu\nu ,\lambda_{1}....\lambda_{s}}$
are used to construct two infinite series of gauge invariant quadratic forms
\cite{Savvidy:2005zm,Savvidy:2005fi}
$$
{{\cal L}}_{s}~~~~~~~s=1,2,3...$$
and \cite{Savvidy:2005fi,Savvidy:2005ki,Savvidy:2005vm}
$${{\cal L}}^{'}_{s}~~~~~~~s=2,3,...$$
Each term of these
infinite series is separately gauge invariant with respect to the generalized gauge
transformations (\ref{polygauge}). These forms
contain quadratic kinetic terms  and nonlinear terms  describing
nonlinear interaction of the Yang-Mills type.
In order to make all tensor gauge fields dynamical one should add
all these forms together. Thus the gauge invariant
Lagrangian describing dynamical tensor gauge bosons of all ranks
has the form
\be\label{generalgaugedensity}
{{\cal L}} = \sum^{\infty}_{s=1}~ g_{s} {{\cal L}}_{s}~+
~ \sum^{\infty}_{s=2}g^{'}_{s} {{\cal L}}^{'}_{s}~,
\ee
where ${{\cal L}}_{1} \equiv {{\cal L}}_{YM}$ is the Yang-Mills Lagrangian.

It is important that:  i) {\it the Lagrangian does not
contain higher derivatives of tensor gauge fields
ii) all interactions take place
through the three- and four-particle exchanges with dimensionless
coupling constant g  iii) the complete Lagrangian contains all higher-rank
tensor gauge fields and should not be truncated iv) the invariance with respect to
the extended gauge transformations does not fix the coupling constants $g_{s}$
and $g^{'}_{s}$}.

The coupling constants $g_{s}$ and $g^{'}_{s}$ remain arbitrary because
every term of the sum is separately gauge invariant and the extended gauge symmetry
alone does not fix them. This means that there is a freedom to vary these
constants without breaking the initial gauge symmetry.
The important question
to which we should address ourselves here is the following:
{\it Can we achieve the enhancement of the initial gauge
symmetry properly tuning the coupling constants $g_{s}$ and $g^{'}_{s}$ ?}

Let us consider a simple example: the sum of
two $Z_2$ invariant forms $g x^2 + g^{'} y^2$ exhibits the
$U(1)$ invariance if we choose $g=g^{'}$ so that the initial symmetry is
elevated to a one parameter family of continuous transformations. A less trivial example
is a linear sum of Poincar\'e invariant forms comprising a SUSY invariant Lagrangian.
One can find other examples
of the same phenomena when a linear sum of invariant forms of the initial group
G exhibits a symmetry with respect to a larger group $\CG \supset G $ when the
coefficients are properly tuned. A similar phenomena appears in our system.

Indeed let us consider a linear sum of two gauge invariant forms in (\ref{generalgaugedensity})
$$
g_{2}{{\cal L}}_{2}+ g^{'}_{2}{{\cal L}}^{'}_{2}
$$
which describes the rank-2 tensor gauge field $A^{a}_{\mu\lambda}$. As we have found in
\cite{Savvidy:2005fi,Savvidy:2005ki,Savvidy:2005vm} one can chose the coupling constants $g_{2}$
and $g^{'}_{2}$ so that the sum $g_{2}{{\cal L}}_{2}+ g^{'}_{2}{{\cal L}}^{'}_{2}$
exhibits invariance with respect to
a bigger gauge group\footnote{$c_2 = g^{'}_{2}/g_{2} = 1$.}.  This means that
in addition to full extended gauge
group (\ref{polygauge}), which we had initially, now we have bigger gauge group
with double number of gauge parameters \cite{Savvidy:2005fi,Savvidy:2005ki,Savvidy:2005vm}.
The explicit form of the free field equation
for the rank-2 tensor gauge field is given by equation (\ref{mainequation}).
It was then demonstrated that it
describes propagation of two polarizations of helicity-two massless charged
tensor gauge boson and of the helicity-zero "axion". This result will be
recapitulated in the third section.

Our aim now is to extend this construction to the rank-3 tensor gauge
field. We shall consider the linear sum
$$
g_{3}{{\cal L}}_{3}+ g^{'}_{3}{{\cal L}}^{'}_{3}
$$
and shall demonstrate that for an appropriate choice of the coupling
constants $c_3 = g^{'}_{3}/g_{3} = 4/3$ the system have an
enhanced gauge symmetry. The explicit description of this symmetry
 together with the corresponding free field equation (\ref{freethirdrankequations})
for the rank-3 tensor gauge field will be given
in the forth and fifth sections. Its relation
to the Schwinger equation for the symmetric rank-3 tensor gauge field is discussed
in the last seventh section.

First let us recapitulate the construction
of the general Lagrangian ${{\cal L}}$ in (\ref{generalgaugedensity}).

\section{\it Non-Abelian Tensor Fields}

The gauge fields are defined as rank-$(s+1)$ tensors
\cite{Savvidy:2005zm,Savvidy:2005fi}
$$
A^{a}_{\mu\lambda_1 ... \lambda_{s}}(x),~~~~~s=0,1,2,...
$$
and are totally symmetric with respect to the
indices $  \lambda_1 ... \lambda_{s}  $. {\it A priory} the tensor fields
have no symmetries with
respect to the first index  $\mu$. The index $a$ numerates the generators $L^a$
of the Lie algebra $\breve{g}$ of a {\it compact}\footnote{The algebra $\breve{g}$
possesses an orthogonal
basis in which the structure constants $f^{abc}$ are totally antisymmetric.}
Lie group G.

One can think of these tensor fields as appearing in the
expansion of the extended gauge field $\CA_{\mu}(x)$ over the unite tangent  vector
$e_{\lambda}$
\cite{Savvidy:2005zm,Savvidy:2005fi}:
\be\label{gaugefield}
\CA_{\mu}(x) = \sum^{\infty}_{s=0}~
A^{a}_{\mu\lambda_1 ... \lambda_{s}}(x) ~L^{a}_{\lambda_1... \lambda_s}.
\ee
The gauge field $A^{a}_{\mu\lambda_1 ... \lambda_{s}}$ carries
indices $a,\lambda_1, ..., \lambda_{s}$ labeling the generators of {\it extended current
algebra $\CG$ associated with compact Lie group G.} It has infinite many generators
$L^{a}_{\lambda_1 ... \lambda_{s}} = L^a e_{\lambda_1}...e_{\lambda_s}$ and
the corresponding algebra is given by the commutator \cite{Savvidy:2005ki}
\footnote{See also the alternative expansions in
\cite{yukawa1,wigner,Savvidy:dv,Savvidy:2003fx,
Savvidy:2005fe,Bengtsson:2004cd,Edgren:2005gq} and the algebras based
on diffeomorphisms
group in \cite{Bakas:1991gs,Bakas:1990xu,Bekaert:2005vh,Ivanov:1979ny}.}
\be
[L^{a}_{\lambda_1 ... \lambda_{s}}, L^{b}_{\rho_1 ... \rho_{k}}]=if^{abc}
L^{c}_{\lambda_1 ... \lambda_{s}\rho_1 ... \rho_{k}}.
\ee

{\it The extended non-Abelian gauge transformations of the
tensor gauge fields are defined
by the following equations } \cite{Savvidy:2005zm,Savvidy:2005fi}:
\beqa\label{polygauge}
\delta A^{a}_{\mu} &=& ( \delta^{ab}\partial_{\mu}
+g f^{acb}A^{c}_{\mu})\xi^b ,~~~~~\\
\delta A^{a}_{\mu\nu} &=&  ( \delta^{ab}\partial_{\mu}
+  g f^{acb}A^{c}_{\mu})\xi^{b}_{\nu} + g f^{acb}A^{c}_{\mu\nu}\xi^{b},\nonumber\\
\delta A^{a}_{\mu\nu \lambda}& =&  ( \delta^{ab}\partial_{\mu}
+g f^{acb} A^{c}_{\mu})\xi^{b}_{\nu\lambda} +
g f^{acb}(  A^{c}_{\mu  \nu}\xi^{b}_{\lambda } +
A^{c}_{\mu \lambda }\xi^{b}_{ \nu}+
A^{c}_{\mu\nu\lambda}\xi^{b}),\nn\\
.........&.&............................\nn
\eeqa
where $\xi^{a}_{\lambda_1 ... \lambda_{s}}(x)$ are totally symmetric gauge parameters.
These extended gauge transformations
generate a closed algebraic structure. To see that, one should compute the
commutator of two extended gauge transformations $\delta_{\eta}$ and $\delta_{\xi}$
of parameters $\eta$ and $\xi$.
The commutator of two transformations can be expressed in the form
\cite{Savvidy:2005zm,Savvidy:2005fi}
\be\label{gaugecommutator}
[~\delta_{\eta},\delta_{\xi}]~A_{\mu\lambda_1\lambda_2 ...\lambda_s} ~=~
-i g~ \delta_{\zeta} A_{\mu\lambda_1\lambda_2 ...\lambda_s}
\ee
and is again an extended gauge transformation with the gauge parameters
$\{\zeta\}$ which are given by the matrix commutators
\beqa\label{gaugealgebra}
\zeta&=&[\eta,\xi]\\
\zeta_{\lambda_1}&=&[\eta,\xi_{\lambda_1}] +[\eta_{\lambda_1},\xi]\nn\\
\zeta_{\nu\lambda} &=& [\eta,\xi_{\nu\lambda}] +  [\eta_{\nu},\xi_{\lambda}]
+ [\eta_{\lambda},\xi_{\nu}]+[\eta_{\nu\lambda},\xi],\nn\\
......&.&..........................\nn
\eeqa
{\it The generalized field strengths  are defined as} \cite{Savvidy:2005zm,Savvidy:2005fi}
\beqa\label{fieldstrengthparticular}
G^{a}_{\mu\nu} &=&
\partial_{\mu} A^{a}_{\nu} - \partial_{\nu} A^{a}_{\mu} +
g f^{abc}~A^{b}_{\mu}~A^{c}_{\nu},\\
G^{a}_{\mu\nu,\lambda} &=&
\partial_{\mu} A^{a}_{\nu\lambda} - \partial_{\nu} A^{a}_{\mu\lambda} +
g f^{abc}(~A^{b}_{\mu}~A^{c}_{\nu\lambda} + A^{b}_{\mu\lambda}~A^{c}_{\nu} ~),\nn\\
G^{a}_{\mu\nu,\lambda\rho} &=&
\partial_{\mu} A^{a}_{\nu\lambda\rho} - \partial_{\nu} A^{a}_{\mu\lambda\rho} +
g f^{abc}(~A^{b}_{\mu}~A^{c}_{\nu\lambda\rho} +
 A^{b}_{\mu\lambda}~A^{c}_{\nu\rho}+A^{b}_{\mu\rho}~A^{c}_{\nu\lambda}
 + A^{b}_{\mu\lambda\rho}~A^{c}_{\nu} ~),\nn\\
 ......&.&............................................\nn
\eeqa
and transform homogeneously with respect to the extended
gauge transformations (\ref{polygauge}). The field strength tensors are
antisymmetric in their first two indices and are totally symmetric with respect to the
rest of the indices.

The inhomogeneous extended gauge transformation (\ref{polygauge})
induces the homogeneous gauge
transformation of the corresponding field strength
(\ref{fieldstrengthparticular}) of the form \cite{Savvidy:2005zm,Savvidy:2005fi}
\beqa\label{fieldstrenghparticulartransformation}
\delta G^{a}_{\mu\nu}&=& g f^{abc} G^{b}_{\mu\nu} \xi^c \\
\delta G^{a}_{\mu\nu,\lambda} &=& g f^{abc} (~G^{b}_{\mu\nu,\lambda} \xi^c
+ G^{b}_{\mu\nu} \xi^{c}_{\lambda}~),\nonumber\\
\delta G^{a}_{\mu\nu,\lambda\rho} &=& g f^{abc}
(~G^{b}_{\mu\nu,\lambda\rho} \xi^c
+ G^{b}_{\mu\nu,\lambda} \xi^{c}_{\rho} +
G^{b}_{\mu\nu,\rho} \xi^{c}_{\lambda} +
G^{b}_{\mu\nu} \xi^{c}_{\lambda\rho}~)\nn\\
......&.&..........................,\nn
\eeqa
The field strength tensors are
antisymmetric in their first two indices and are totally symmetric with respect to the
rest of the indices.
The symmetry properties of the field strength  $G^{a}_{\mu\nu,\lambda_1 ... \lambda_s}$
remain invariant in the course of this transformation.

These tensor gauge fields and the corresponding field strength tensors allow to
construct two series of gauge invariant quadratic forms. The
first series is given by the formula \cite{Savvidy:2005zm,Savvidy:2005fi}:
\beqa\label{fulllagrangian1}
{{\cal L}}_{s+1}&=&-{1\over 4} ~
G^{a}_{\mu\nu, \lambda_1 ... \lambda_s}~
G^{a}_{\mu\nu, \lambda_{1}...\lambda_{s}} +.......\nonumber\\
&=& -{1\over 4}\sum^{2s}_{i=0}~a^{s}_i ~
G^{a}_{\mu\nu, \lambda_1 ... \lambda_i}~
G^{a}_{\mu\nu, \lambda_{i+1}...\lambda_{2s}}
(\sum_{p's} \eta^{\lambda_{i_1} \lambda_{i_2}} .......
\eta^{\lambda_{i_{2s-1}} \lambda_{i_{2s}}})~,
\eeqa
where the sum $\sum_p$ runs over all nonequal permutations of $i's$, in total $(2s-1)!!$
terms and the numerical coefficients are $a^{s}_i = {s!\over i!(2s-i)!}$.

The second series of gauge invariant quadratic forms is given by the formula
\cite{Savvidy:2005fi,Savvidy:2005ki,Savvidy:2005vm}:
\beqa\label{secondfulllagrangian}
{{\cal L}}^{'}_{s+1}&=&{1\over 4} ~
G^{a}_{\mu\lambda_1,\lambda_2  ... \lambda_{s+1}}~
G^{a}_{\mu\lambda_2,\lambda_{1} ...\lambda_{s+1}} +.......\nonumber\\
&=& {1\over 8}\sum^{2s+1}_{i=1}~a^{s}_{i-1} ~
G^{a}_{\mu\lambda_1,\lambda_2  ... \lambda_i}~
G^{a}_{\mu\lambda_{i+1},\lambda_{i+2} ...\lambda_{2s+2}}
(\sum^{'}_{p's} \eta^{\lambda_{i_1} \lambda_{i_2}} .......
\eta^{\lambda_{i_{2s+1}} \lambda_{i_{2s+2}}})~,
\eeqa
where the sum $\sum^{'}_p$ runs over all nonequal permutations of $i's$, with exclusion
of the terms which contain $\eta^{\lambda_{1},\lambda_{i+1}}$.

In order to make all tensor gauge fields dynamical one should add
the corresponding kinetic terms. Thus the invariant
Lagrangian describing dynamical tensor gauge bosons of all ranks
has the form
\be\label{fulllagrangian2}
{{\cal L}} = \sum^{\infty}_{s=1}~ g_{s } {{\cal L}}_{s }~+
\sum^{\infty}_{s=2}~ g^{'}_{s } {{\cal L}}^{'}_{s }~,
\ee
where ${{\cal L}}_{1} \equiv {{\cal L}}_{YM}$.

It is important that:  i)  the Lagrangian does not
contain higher derivatives of tensor gauge fields
ii) all interactions take place
through the three- and four-particle exchanges with dimensionless
coupling constant g iii) the complete Lagrangian contains all higher-rank
tensor gauge fields and should not be truncated iv) the invariance with respect to
the extended gauge transformations
does not fix the coupling constants $g_{s}$
and $g^{'}_{s}$.

The coupling constants $g_{s}$ and $g^{'}_{s}$ remain arbitrary because
every term of the sum is separately gauge invariant and the extended gauge symmetry
alone does not fix them. This means that we have a freedom to chose these
constants without breaking the initial gauge symmetry.
The important question
which should be addressed here is the following: {\it Can we achieve the enhancement of
the gauge symmetry  tuning the coupling constants $g_{s}$ and $g^{'}_{s}$ ?}

Let us consider a linear sum of two gauge invariant forms in (\ref{generalgaugedensity})
$$
g_{2}{{\cal L}}_{2}+ g^{'}_{2}{{\cal L}}^{'}_{2}~,
$$
which describes the rank-2 tensor gauge field $A^{a}_{\mu\lambda}$. As we have found in
\cite{Savvidy:2005fi,Savvidy:2005ki,Savvidy:2005vm}
one can choose the coupling constants $g_{2}$
and $g^{'}_{2}$ so that the sum $g_{2}{{\cal L}}_{2}+ g^{'}_{2}{{\cal L}}^{'}_{2}$
exhibits invariance with respect to
a bigger gauge group (see the next section for details, where we will
demonstrate that $c_2 = g^{'}_{2}/g_{2} = 1$).
This means that in addition to full extended gauge
group (\ref{polygauge}), which we had initially, now we have a bigger gauge group
with double number of gauge parameters \cite{Savvidy:2005fi,Savvidy:2005ki,Savvidy:2005vm}.
It was then demonstrated that the corresponding kinetic term
describes propagation of two polarizations of helicity-two massless charged
tensor gauge boson and the helicity-zero "axion".

In summary the gauge invariant Lagrangian for the lower-rank tensor
gauge fields has the form \cite{Savvidy:2005fi,Savvidy:2005ki,Savvidy:2005vm}:
\beqa\label{totalactiontwo}
{{\cal L}}=  {{\cal L}}_1 +   {{\cal L}}_2 +  {{\cal L}}^{'}_2  =
&-&{1\over 4}G^{a}_{\mu\nu}G^{a}_{\mu\nu}\\
&-&{1\over 4}G^{a}_{\mu\nu,\lambda}G^{a}_{\mu\nu,\lambda}
-{1\over 4}G^{a}_{\mu\nu}G^{a}_{\mu\nu,\lambda\lambda}\nn\\
&+&{1\over 4}G^{a}_{\mu\nu,\lambda}G^{a}_{\mu\lambda,\nu}
+{1\over 4}G^{a}_{\mu\nu,\nu}G^{a}_{\mu\lambda,\lambda}
+{1\over 2}G^{a}_{\mu\nu}G^{a}_{\mu\lambda,\nu\lambda}.\nn
\eeqa
The equations of motion which follow from this Lagrangian are:
\beqa\label{equationforfirstranktensor}
\nabla^{ab}_{\mu}G^{b}_{\mu\nu}
&+&{1\over 2 }\nabla^{ab}_{\mu} (G^{b}_{\mu\nu,\lambda\lambda}
+ G^{b}_{\nu\lambda,\mu\lambda}
+ G^{b}_{\lambda\mu,\nu\lambda})
+ g f^{acb} A^{c}_{\mu\lambda} G^{b}_{\mu\nu,\lambda}\\
&-&{1\over 2 }g f^{acb} (A^{c}_{\mu\lambda} G^{b}_{\mu\lambda,\nu}
+A^{c}_{\mu\lambda} G^{b}_{\lambda\nu,\mu}
+A^{c}_{\lambda\lambda} G^{b}_{\mu\nu,\mu}
+A^{c}_{\mu\nu} G^{b}_{\mu\lambda,\lambda})\nn\\
&+&{1\over 2 }g f^{acb} (
A^{c}_{\mu\lambda\lambda} G^{b}_{\mu\nu}
+ A^{c}_{\mu \mu\lambda } G^{b}_{\nu \lambda}
+ A^{c}_{\mu\nu\lambda} G^{b}_{\lambda\mu})\nn
=0
\eeqa\label{equationforsecondranktensor}
and for the second-rank tensor gauge field $A^{a}_{\nu\lambda}$:
\beqa\label{secondrankfieldequations}
&&\nabla^{ab}_{\mu}G^{b}_{\mu\nu,\lambda}
-{1\over 2} (\nabla^{ab}_{\mu}G^{b}_{\mu\lambda,\nu}
+\nabla^{ab}_{\mu}G^{b}_{\lambda\nu,\mu}
+\nabla^{ab}_{\lambda}G^{b}_{\mu\nu,\mu}
+\eta_{\nu\lambda} \nabla^{ab}_{\mu}G^{b}_{\mu\rho,\rho})\nn\\
&+&g f^{acb} A^{c}_{\mu\lambda} G^{b}_{\mu\nu} -
{1\over 2}g f^{acb}(A^{c}_{\mu\nu} G^{b}_{\mu\lambda}
+A^{c}_{\mu\mu} G^{b}_{\lambda\nu}
+A^{c}_{\lambda\mu} G^{b}_{\mu\nu}
+\eta_{\nu\lambda}  A^{c}_{\mu\rho} G^{b}_{\mu\rho})
=0.
\eeqa
The variation of the action with respect to the third-rank gauge field
$A^{a}_{\nu\lambda\rho}$ will give the equations
\be
\eta_{\lambda\rho}\nabla^{ab}_{\mu}G^{b}_{\mu\nu}-{1\over 2}
(\eta_{\nu\rho}\nabla^{ab}_{\mu}G^{b}_{\mu\lambda}  +
\eta_{\lambda\nu}\nabla^{ab}_{\mu}G^{b}_{\mu\rho}) +
{1\over 2} (\nabla^{ab}_{\rho}G^{b}_{\nu\lambda}  +
\nabla^{ab}_{\lambda}G^{b}_{\nu\rho})=0.
\ee
Representing this system of equations in the form
\beqa\label{perturbativeform}
\partial_{\mu} F^{a}_{\mu\nu}
+{1\over 2 }\partial_{\mu} (F^{a}_{\mu\nu,\lambda\lambda}
+ F^{a}_{\nu\lambda,\mu\lambda}
+ F^{a}_{\lambda\mu,\nu\lambda})= j^{a}_{\nu}\\
\partial_{\mu} F^{a}_{\mu\nu,\lambda}
-{1\over 2} (\partial_{\mu} F^{a}_{\mu\lambda,\nu}
+\partial_{\mu} F^{a}_{\lambda\nu,\mu}
+\partial_{\lambda}F^{a}_{\mu\nu,\mu}
+\eta_{\nu\lambda} \partial_{\mu}F^{a}_{\mu\rho,\rho}) = j^{a}_{\nu\lambda}\nn\\
\eta_{\lambda\rho}\partial_{\mu}F^{a}_{\mu\nu}-{1\over 2}
(\eta_{\nu\rho}\partial_{\mu}F^{a}_{\mu\lambda}  +
\eta_{\nu\lambda}\partial_{\mu}F^{a}_{\mu\rho}) +
{1\over 2} (\partial_{\rho}F^{a}_{\nu\lambda}  +
\partial_{\lambda}F^{a}_{\nu\rho})=
j^{a}_{\nu\lambda\rho},\nn
\eeqa
where $F^{a}_{\mu\nu} =  \partial_{\mu} A^{a}_{\nu  } -
\partial_{\nu} A^{a}_{\mu },~
F^{a}_{\mu\nu,\lambda} = \partial_{\mu} A^{a}_{\nu \lambda} -
\partial_{\nu} A^{a}_{\mu \lambda},~
F^{a}_{\mu\nu,\lambda\rho} = \partial_{\mu} A^{a}_{\nu \lambda\rho} -
\partial_{\nu} A^{a}_{\mu \lambda\rho}$ ,
we can find the corresponding conserved currents
\beqa\label{vectorcurrent}
j^{a}_{\nu } = &-&g f^{abc} A^{b}_{\mu } G^{c}_{\mu\nu }
-g f^{abc}\partial_{\mu} (A^{b}_{\mu } A^{c}_{\nu })\\
&-&{1\over 2 }g f^{abc}A^{b}_{\mu} (G^{c}_{\mu\nu,\lambda\lambda}
+ G^{c}_{\nu\lambda,\mu\lambda}
+ G^{c}_{\lambda\mu,\nu\lambda})
-{1\over 2 }\partial_{\mu} (I^{a}_{\mu\nu,\lambda\lambda}
+ I^{a}_{\nu\lambda,\mu\lambda}
+ I^{a}_{\lambda\mu,\nu\lambda})\nn\\
&-& g f^{abc} A^{b}_{\mu\lambda} G^{c}_{\mu\nu,\lambda}
+ {1\over 2 }g f^{abc} (A^{b}_{\mu\lambda} G^{c}_{\mu\lambda,\nu}
+A^{b}_{\mu\lambda} G^{c}_{\lambda\nu,\mu}
+A^{b}_{\lambda\lambda} G^{c}_{\mu\nu,\mu}
+A^{b}_{\mu\nu} G^{c}_{\mu\lambda,\lambda}) \nn\\
&-&{1\over 2} g f^{abc} (A^{b}_{\mu\lambda\lambda} G^{c}_{\mu\nu}
+A^{b}_{\lambda\mu\lambda} G^{c}_{\nu\mu}
+ A^{b}_{\mu\lambda\nu} G^{c}_{\lambda\mu}),\nn
\eeqa
where $I^{a}_{\mu\nu,\lambda\rho}=g f^{abc}(~A^{b}_{\mu}~A^{c}_{\nu\lambda\rho} +
 A^{b}_{\mu\lambda}~A^{c}_{\nu\rho}+A^{b}_{\mu\rho}~A^{c}_{\nu\lambda}
 + A^{b}_{\mu\lambda\rho}~A^{c}_{\nu} ~)$ and
\beqa\label{tensorcurrent}
j^{a}_{\nu\lambda}=&-&g f^{abc} A^{b}_{\mu} G^{c}_{\mu\nu,\lambda}
+{1\over 2 }g f^{abc} (A^{b}_{\mu} G^{c}_{\mu\lambda,\nu}
+A^{b}_{\mu} G^{c}_{\lambda\nu,\mu}
+A^{b}_{\lambda} G^{c}_{\mu\nu,\mu}
+\eta_{\nu\lambda}A^{b}_{\mu} G^{c}_{\mu\rho,\rho})\nn\\
&-&g f^{abc} A^{b}_{\mu\lambda} G^{c}_{\mu\nu} +
{1\over 2}g f^{abc}(A^{b}_{\mu\nu} G^{c}_{\mu\lambda}
+A^{b}_{\lambda\mu} G^{c}_{\mu\nu}
+A^{b}_{\mu\mu} G^{c}_{\lambda\nu}
+\eta_{\nu\lambda}  A^{b}_{\mu\rho} G^{c}_{\mu\rho})\nn\\
&-&g f^{abc} \partial_{\mu}
(A^{b}_{\mu} A^{c}_{\nu\lambda} + A^{b}_{\mu\lambda} A^{c}_{\nu}) +
{1\over 2}g f^{abc}
[\partial_{\mu}(A^{b}_{\mu} A^{c}_{\lambda\nu}+A^{b}_{\mu\nu} A^{c}_{\lambda})
+\partial_{\mu}(A^{b}_{\lambda} A^{c}_{\nu\mu}+A^{b}_{\lambda\mu} A^{c}_{\nu})\nn\\
&+&\partial_{\lambda} (A^{b}_{\mu} A^{b}_{\nu\mu}  + A^{b}_{\mu\mu} A^{c}_{\nu})
+\eta_{\nu\lambda}  \partial_{\mu}
(A^{b}_{\mu} A^{b}_{\rho\rho} + A^{b}_{\mu\rho} A^{c}_{\rho})],
\eeqa
\beqa\label{tensorcurrentthierd}
j^{a}_{\nu\lambda\rho}=&-&\eta_{\lambda\rho} ~g f^{abc} A^{b}_{\mu} G^{c}_{\mu\nu}
+{1\over 2 }g f^{abc} (\eta_{\nu\rho} A^{b}_{\mu} G^{c}_{\mu\lambda}
+\eta_{\nu\lambda}A^{b}_{\mu} G^{c}_{\mu\rho}
-A^{b}_{\rho} G^{c}_{\nu\lambda}
-A^{b}_{\lambda} G^{c}_{\nu\rho})\\
&-&\eta_{\lambda\rho} ~g f^{abc} \partial_{\mu}
(A^{b}_{\mu} A^{c}_{\nu}) +
{1\over 2}g f^{abc}
[\partial_{\mu}(\eta_{\nu\lambda} A^{b}_{\mu} A^{c}_{\rho}
+ \eta_{\nu\rho} A^{b}_{\mu} A^{c}_{\lambda})
-\partial_{\lambda} (A^{b}_{\nu} A^{c}_{\rho})
-\partial_{\rho}(A^{b}_{\nu} A^{c}_{\lambda})].\nn
\eeqa
The conservation of the corresponding currents follows from the fact that
we have enhancement of the gauge group and therefore
the partial derivatives of the l.h.s. of the equations
(\ref{perturbativeform}) are equal to zero
\beqa\label{gaugecurrentconservation}
\partial_{\nu} j^{a}_{\nu}&=&0,~~~\nn\\
\partial_{\nu} j^{a}_{\nu\lambda}&=&0,~~~~
\partial_{\lambda} j^{a}_{\nu\lambda}=0,\nn\\
\partial_{\nu} j^{a}_{\nu\lambda\rho}&=&0,~~~~
\partial_{\lambda} j^{a}_{\nu\lambda\rho}=0,~~~~
\partial_{\rho} j^{a}_{\nu\lambda\rho}=0.
\eeqa

Our aim now is to extend this analysis to the case of rank-3 tensor gauge
field. We shall consider the linear sum
$$
g_{3}{{\cal L}}_{3}+ g^{'}_{3}{{\cal L}}^{'}_{3}
$$
and demonstrate that for an appropriate choice of the coupling
constants ratio $c_3 = g^{'}_{3}/g_{3} = 4/3$ the system will have an
enhanced symmetry.
The explicit form of the Lagrangian for the third-rank tensor gauge field
$A^{a}_{\mu\nu\lambda}$ can be obtained from our general
formulas (\ref{fulllagrangian1}), (\ref{secondfulllagrangian}) and
(\ref{fulllagrangian2}) when we substitute
$s=2$.
The Lagrangian is:
\beqa\label{thirdranktensorlagrangian}
{{\cal L}}_3 + c_3{{\cal L}}^{'}_3
=&-&{1\over 4}G^{a}_{\mu\nu,\lambda\rho}G^{a}_{\mu\nu,\lambda\rho}
-{1\over 8}G^{a}_{\mu\nu ,\lambda\lambda}G^{a}_{\mu\nu ,\rho\rho}
-{1\over 2}G^{a}_{\mu\nu,\lambda}  G^{a}_{\mu\nu ,\lambda \rho\rho}
-{1\over 8}G^{a}_{\mu\nu}  G^{a}_{\mu\nu ,\lambda \lambda\rho\rho}+ \nn\\
&+&c_3\{{1\over 4}
G^{a}_{\mu\nu,\lambda\rho}G^{a}_{\mu\lambda,\nu\rho}+
{1\over 4} G^{a}_{\mu\nu,\nu\lambda}G^{a}_{\mu\rho,\rho\lambda}+
{1\over 4}G^{a}_{\mu\nu,\nu\lambda}G^{a}_{\mu\lambda,\rho\rho}+\\
&+&{1\over 4}G^{a}_{\mu\nu,\lambda}G^{a}_{\mu\lambda,\nu\rho\rho}
+{1\over 2}G^{a}_{\mu\nu,\lambda}G^{a}_{\mu\rho,\nu\lambda\rho}
+{1\over 4}G^{a}_{\mu\nu,\nu}G^{a}_{\mu\lambda,\lambda\rho\rho}
+{1\over 4}G^{a}_{\mu\nu}G^{a}_{\mu\lambda,\nu\lambda\rho\rho}\},\nn
\eeqa
where $c_3 = g^{'}_{3}/g_{3}$ is a constant\footnote{It is not difficult to present the
explicit form of the Lagrangian for any higher-rank tensor field using
expressions (\ref{fulllagrangian1}), (\ref{secondfulllagrangian})
and (\ref{fulllagrangian2}).}.

\section{\it Enhanced Symmetry. Rank-2 Gauge Field}

As we have seen above there are two invariant forms for the rank-2 tensor gauge
field ${{\cal L}}_2$ and ${{\cal L}}^{'}_2$ and the general Lagrangian is a
linear combination
$
{{\cal L}}_2 + c_2 {{\cal L}}^{'}_2 ,
$
where $c_2 = g^{'}_{2}/g_2 $ is a constant coefficient. Let us review how
this coefficient has been fixed
by the requirement of an enhanced symmetry.
For that let us consider the
situation at the linearized level when the gauge coupling constant g is equal to zero.
The free part of the ${{\cal L}}_2$ Lagrangian
$$
{{\cal L}}^{free}_2 ={1 \over 2} A^{a}_{\alpha\acute{\alpha}}
(\eta_{\alpha\gamma}\eta_{\acute{\alpha}\acute{\gamma}}\partial^{2} -
\eta_{\acute{\alpha}\acute{\gamma}} \partial_{\alpha} \partial_{\gamma} )
A^{a}_{\gamma\acute{\gamma}} =
{1 \over 2} A^{a}_{\alpha\acute{\alpha}}
H_{\alpha\acute{\alpha}\gamma\acute{\gamma}} A^{a}_{\gamma\acute{\gamma}} ,
$$
where the quadratic form in the momentum representation has the form
$$
H_{\alpha\acute{\alpha}\gamma\acute{\gamma}}(k)=
(-k^2 \eta_{\alpha\gamma} +k_{\alpha}k_{\gamma})
\eta_{\acute{\alpha}\acute{\gamma}},
$$
is obviously invariant with respect to the gauge
transformation $\delta A^{a}_{\mu\lambda} =\partial_{\mu} \xi^{a}_{\lambda}$,
but it is not invariant with respect to the alternative gauge transformations
$\tilde{\delta} A^{a}_{\mu \lambda} =\partial_{\lambda} \eta^{a}_{\mu}$. This can be
seen, for example, from the following relations in momentum representation
\be\label{currentdivergence}
k_{\alpha}H_{\alpha\acute{\alpha}\gamma\acute{\gamma}}(k)=0,~~~
k_{\acute{\alpha}}H_{\alpha\acute{\alpha}\gamma\acute{\gamma}}(k)=
-(k^2 \eta_{\alpha\gamma} - k_{\alpha}k_{\gamma})k_{\acute{\gamma}} \neq 0 .
\ee
Let us consider now the free part of the second Lagrangian
\beqa
{{\cal L}}^{' free}_{2} ={1 \over 4} A^{a}_{\alpha\acute{\alpha}}
(-\eta_{\alpha\acute{\gamma}}\eta_{\acute{\alpha}\gamma}\partial^{2} -
\eta_{\alpha\acute{\alpha}}\eta_{\gamma\acute{\gamma}}\partial^{2}
+\eta_{\alpha\acute{\gamma}} \partial_{\acute{\alpha}} \partial_{\gamma}
+\eta_{\acute{\alpha}\gamma} \partial_{\alpha} \partial_{\acute{\gamma}}
+\eta_{\alpha\acute{\alpha}} \partial_{\gamma} \partial_{\acute{\gamma}}+\nn\\
+\eta_{\gamma\acute{\gamma}} \partial_{\alpha} \partial_{\acute{\alpha}}
-2\eta_{\alpha\gamma} \partial_{\acute{\alpha}} \partial_{\acute{\gamma}})
A^{a}_{\gamma\acute{\gamma}}=
{1 \over 2} A^{a}_{\alpha\acute{\alpha}}
H^{~'}_{\alpha\acute{\alpha}\gamma\acute{\gamma}} A^{a}_{\gamma\acute{\gamma}},
\eeqa
where
$$
H^{'}_{\alpha\acute{\alpha}\gamma\acute{\gamma}}(k)=
{1 \over 2}(\eta_{\alpha\acute{\gamma}}\eta_{\acute{\alpha}\gamma}
+\eta_{\alpha\acute{\alpha}}\eta_{\gamma\acute{\gamma}})k^2
-{1 \over 2}(\eta_{\alpha\acute{\gamma}}k_{\acute\alpha}k_{\gamma}
+\eta_{\acute\alpha\gamma}k_{\alpha}k_{\acute{\gamma}}
+\eta_{\alpha\acute\alpha}k_{\gamma}k_{\acute{\gamma}}
+\eta_{\gamma\acute{\gamma}}k_{\alpha}k_{\acute\alpha}
-2\eta_{\alpha\gamma}k_{\acute\alpha}k_{\acute{\gamma}}).
$$
It is again invariant with respect to the gauge
transformation $\delta A^{a}_{\mu\lambda} =\partial_{\mu} \xi^{a}_{\lambda}$,
but it is not invariant with respect to the gauge transformations
$\tilde{\delta} A^{a}_{\mu \lambda} =\partial_{\lambda} \eta^{a}_{\mu}$, as one can
see from analogous relations
\be\label{currentdivergenceprime}
k_{\alpha}H^{'}_{\alpha\acute{\alpha}\gamma\acute{\gamma}}(k)=0,~~~
k_{\acute{\alpha}}H^{'}_{\alpha\acute{\alpha}\gamma\acute{\gamma}}(k)=
(k^2 \eta_{\alpha\gamma} -k_{\alpha}k_{\gamma})k_{\acute{\gamma}} \neq 0 .
\ee
As it is obvious from (\ref{currentdivergence}) and
(\ref{currentdivergenceprime}), the total Lagrangian
${{\cal L}}^{free}_2 + {{\cal L}}^{' free}_2$ now  poses new enhanced
invariance with respect to the larger, eight-parameter, gauge transformations
\be\label{largegaugetransformation}
\delta A^{a}_{\mu \lambda} =\partial_{\mu} \xi^{a}_{\lambda}+
\partial_{\lambda} \eta^{a}_{\mu} ,
\ee
where $\xi^{a}_{\lambda}$ and $\eta^{a}_{\mu}$ are eight arbitrary functions, because
\be\label{zeroderivatives}
k_{\alpha}(H_{\alpha\acute{\alpha}\gamma\acute{\gamma}}+
H^{'}_{\alpha\acute{\alpha}\gamma\acute{\gamma}})=0,~~~
k_{\acute{\alpha}}(H_{\alpha\acute{\alpha}\gamma\acute{\gamma}}+
H^{'}_{\alpha\acute{\alpha}\gamma\acute{\gamma}})=0 .
\ee
Thus our free part of the Lagrangian is
\beqa\label{totalfreelagrangian}
{{\cal L}}^{tot~free}_{2} =&-&{1 \over 2}\partial_{\mu}
A^{a}_{\nu \lambda}\partial_{\mu} A^{a}_{\nu \lambda}
+{1 \over 2}\partial_{\mu} A^{a}_{\nu \lambda}\partial_{\nu} A^{a}_{\mu \lambda}+
\nn\\
&+&{1 \over 4} \partial_{\mu} A^{a}_{\nu \lambda} \partial_{\mu } A^{a}_{\lambda\nu}
-{1 \over 4} \partial_{\mu} A^{a}_{\nu \lambda} \partial_{\lambda} A^{a}_{\mu \nu}
-{1 \over 4}\partial_{\nu} A^{a}_{\mu \lambda} \partial_{\mu} A^{a}_{\lambda\nu }
+{1 \over 4} \partial_{\nu } A^{a}_{\mu\lambda} \partial_{\lambda} A^{a}_{\mu \nu}
\nn\\
&+&{1 \over 4}\partial_{\mu} A^{a}_{\nu \nu}\partial_{\mu} A^{a}_{\lambda\lambda}
-{1 \over 2}\partial_{\mu} A^{a}_{\nu \nu} \partial_{\lambda} A^{a}_{\mu\lambda}
+{1 \over 4}\partial_{\nu } A^{a}_{\mu\nu}\partial_{\lambda} A^{a}_{\mu\lambda}
\eeqa
or, in equivalent form, it is
\beqa\label{totfreelagrangianalternativeform}
{{\cal L}}^{tot~free}_{2} ={1 \over 2} A^{a}_{\alpha\acute{\alpha}}
\{(\eta_{\alpha\gamma}\eta_{\acute{\alpha}\acute{\gamma}}
-{1\over 2}\eta_{\alpha\acute{\gamma}}\eta_{\acute{\alpha}\gamma}
-{1\over 2}\eta_{\alpha\acute{\alpha}}\eta_{\gamma\acute{\gamma}})
\partial^{2}
-\eta_{\acute{\alpha}\acute{\gamma}} \partial_{\alpha} \partial_{\gamma}
-\eta_{\alpha\gamma} \partial_{\acute{\alpha}} \partial_{\acute{\gamma}}+\nn\\
+{1\over 2}(\eta_{\alpha\acute{\gamma}} \partial_{\acute{\alpha}} \partial_{\gamma}
+\eta_{\acute{\alpha}\gamma} \partial_{\alpha} \partial_{\acute{\gamma}}
+\eta_{\alpha\acute{\alpha}} \partial_{\gamma} \partial_{\acute{\gamma}}
+\eta_{\gamma\acute{\gamma}} \partial_{\alpha} \partial_{\acute{\alpha}})
\}
A^{a}_{\gamma\acute{\gamma}}
\eeqa
and is invariant with respect to the larger gauge transformations
$
\delta A^{a}_{\mu \lambda} =\partial_{\mu} \xi^{a}_{\lambda}+
\partial_{\lambda} \eta^{a}_{\mu},
$
where $\xi^{a}_{\lambda}$ and $\eta^{a}_{\mu}$ are eight arbitrary functions.
In momentum representation the quadratic form is
\beqa\label{quadraticform}
H^{tot}_{\alpha\acute{\alpha}\gamma\acute{\gamma}}(k)=
(-\eta_{\alpha\gamma}\eta_{\acute{\alpha}\acute{\gamma}}
+{1 \over 2}\eta_{\alpha\acute{\gamma}}\eta_{\acute{\alpha}\gamma}
+{1 \over 2}\eta_{\alpha\acute{\alpha}}\eta_{\gamma\acute{\gamma}})k^2
+\eta_{\alpha\gamma}k_{\acute\alpha}k_{\acute{\gamma}}
+\eta_{\acute\alpha \acute{\gamma}}k_{\alpha}k_{\gamma}\nn\\
-{1 \over 2}(\eta_{\alpha\acute{\gamma}}k_{\acute\alpha}k_{\gamma}
+\eta_{\acute\alpha\gamma}k_{\alpha}k_{\acute{\gamma}}
+\eta_{\alpha\acute\alpha}k_{\gamma}k_{\acute{\gamma}}
+\eta_{\gamma\acute{\gamma}}k_{\alpha}k_{\acute\alpha}).
\eeqa
Free equations of motion which follow from the Lagrangian
(\ref{totfreelagrangianalternativeform}) will take the form
\beqa\label{mainequation}
\partial^{2}(A^{a}_{\nu\lambda} -{1\over 2}A^{a}_{\lambda\nu})
-\partial_{\nu} \partial_{\mu}  (A^{a}_{\mu\lambda}-
{1\over 2}A^{a}_{\lambda\mu} )&-&
\partial_{\lambda} \partial_{\mu}  (A^{a}_{\nu\mu} - {1\over 2}A^{a}_{\mu\nu} )
+\partial_{\nu} \partial_{\lambda} ( A^{a}_{\mu\mu}-{1\over 2}A^{a}_{\mu\mu})\nn\\
&+&{1\over 2}\eta_{\nu\lambda} ( \partial_{\mu} \partial_{\rho}A^{a}_{\mu\rho}
-  \partial^{2}A^{a}_{\mu\mu})=0
\eeqa
and  describe the propagation of massless particles
of spin 2 and spin 0. It is also easy to see that for the symmetric
tensor gauge fields
$A^{a}_{\nu\lambda} = A^{a}_{\lambda\nu}$ our equation
reduces to the  Einstein-Fierz-Pauli-Schwinger-Chang-Singh-Hagen-Fronsdal equation
$$
\partial^{2} A_{\nu\lambda}
-\partial_{\nu} \partial_{\mu}  A_{\mu\lambda} -
\partial_{\lambda} \partial_{\mu}  A_{\mu\nu}
+ \partial_{\nu} \partial_{\lambda}  A_{\mu\mu}
+\eta_{\nu\lambda}  (\partial_{\mu} \partial_{\rho}A_{\mu\rho}
- \partial^{2} A_{\mu\mu}) =0,
$$
which describes the propagation of massless boson with two
physical polarizations, the $s= \pm 2$ helicity states.
For the antisymmetric fields it reduces to the equation
$$
\partial^{2} A_{\nu\lambda}
-\partial_{\nu} \partial_{\mu}  A_{\mu\lambda} +
\partial_{\lambda} \partial_{\mu}  A_{\mu\nu} =0,
$$
and describes the propagation of one physical polarization $s= 0$,
the zero helicity state.

The above consideration brings the final form of the gauge invariant Lagrangian
for the lower-rank tensor gauge fields to the form (\ref{totalactiontwo}).

\section{\it Rank-3 Tensor Gauge Field}

The Lagrangian $ {{\cal L}}_1 +  g_2({{\cal L}}_2 +  {{\cal L}}^{'}_2)$ contains the
third-rank gauge fields $A^{a}_{\mu\nu\lambda}$, but without corresponding
kinetic term. In order to make the fields $A^{a}_{\mu\nu\lambda}$ dynamical
we have to add the corresponding Lagrangian
$g_{3}{{\cal L}}_{3}+ g^{'}_{3}{{\cal L}}^{'}_{3}$
presented in  (\ref{thirdranktensorlagrangian}),
so that at this level the total Lagrangian is the sum
\cite{Savvidy:2005fi,Savvidy:2005ki,Savvidy:2005vm}
$$
{{\cal L}}=
{{\cal L}}_1 +g_2 ({{\cal L}}_2 +{{\cal L}}^{'}_2 )
+g_3( {{\cal L}}_3 +c_3 {{\cal L}}^{'}_3 )~+... ,
$$
where $c_3 = g^{'}_3/g_3$. The Lagrangian ${{\cal L}}_3$ has
the form (\ref{thirdranktensorlagrangian}):
\beqa
{{\cal L}}_3 =-{1\over 4}G^{a}_{\mu\nu,\lambda\rho}G^{a}_{\mu\nu,\lambda\rho}
-{1\over 8}G^{a}_{\mu\nu ,\lambda\lambda}G^{a}_{\mu\nu ,\rho\rho}
-{1\over 2}G^{a}_{\mu\nu,\lambda}  G^{a}_{\mu\nu ,\lambda \rho\rho}
-{1\over 8}G^{a}_{\mu\nu}  G^{a}_{\mu\nu ,\lambda \lambda\rho\rho}~,
\eeqa
where the field strength tensors (\ref{fieldstrengthparticular}) are
\beqa\label{spin4fieldstrenghth}
G^{a}_{\mu\nu ,\lambda \rho \sigma} =
\partial_{\mu} A^{a}_{\nu \lambda \rho \sigma} -
\partial_{\nu} A^{a}_{\mu \lambda\rho\sigma} +
g f^{abc}\{~A^{b}_{\mu}~A^{c}_{\nu \lambda \rho\sigma}
+A^{b}_{\mu\lambda}~A^{c}_{\nu\rho \sigma} +
A^{b}_{\mu\rho }~A^{c}_{\nu\lambda\sigma} +
A^{b}_{\mu\sigma}~A^{c}_{\nu\lambda\rho} +\nn\\
+A^{b}_{\mu\lambda\rho}~A^{c}_{\nu \sigma} +
A^{b}_{\mu\lambda\sigma}~A^{c}_{\nu\rho} +
A^{b}_{\mu\rho\sigma}~A^{c}_{\nu \lambda} +
     A^{b}_{\mu\lambda\rho\sigma }~A^{c}_{\nu} ~\}\nonumber
\eeqa
and
\beqa\label{spin4fieldstrenghth4}
G^{a}_{\mu\nu ,\lambda \rho \sigma\delta} =
\partial_{\mu} A^{a}_{\nu \lambda \rho \sigma\delta} -
\partial_{\nu} A^{a}_{\mu \lambda\rho\sigma\delta} &+&
g f^{abc}\{~A^{b}_{\mu}~A^{c}_{\nu \lambda \rho\sigma\delta}
+\sum_{ \lambda \leftrightarrow \rho,\sigma,\delta}
        A^{b}_{\mu\lambda}~A^{c}_{\nu\rho \sigma\delta} + \nn\\
&+&\sum_{\lambda,\rho \leftrightarrow \sigma,\delta}
        A^{b}_{\mu\lambda\rho}~A^{c}_{\nu\sigma\delta} +
\sum_{\lambda,\rho,\sigma\leftrightarrow \delta}
       A^{b}_{\mu\lambda\rho\sigma}~A^{c}_{\nu\delta} +
     A^{b}_{\mu\lambda\rho\sigma\delta }~A^{c}_{\nu} ~\}.\nonumber
\eeqa
The terms in parentheses are symmetric over $\lambda \rho\sigma$ and
$\lambda \rho \sigma\delta$ respectively. The Lagrangian ${{\cal L}}_3$
is invariant with respect to the extended gauge transformations  (\ref{polygauge})
of the low-rank gauge fields
$ A_{\mu}, A_{\mu\nu}, A_{\mu\nu\lambda}$  and of the fourth-rank gauge field
(\ref{polygauge})
\beqa\label{gaugetransform4}
\delta_{\xi}  A_{\mu\nu\lambda\rho} =\partial_{\mu}\xi_{\nu\lambda\rho}
-i g[A_{\mu},\xi_{\nu\lambda\rho}]
-i g [A_{\mu\nu},\xi_{\lambda\rho}]
-i g [A_{\mu\lambda},\xi_{\nu\rho}]
-i g [A_{\mu\rho},\xi_{\nu\lambda}]-\nn\\
-i g  [A_{\mu\nu\lambda},\xi_{\rho}]
-i g  [A_{\mu\nu\rho},\xi_{\lambda}]
-i g  [A_{\mu\lambda\rho},\xi_{\nu}]
-i g [A_{\mu\nu\lambda\rho},\xi]\nonumber
\eeqa
and of the fifth-rank tensor gauge field (\ref{polygauge})
\beqa\label{gaugetransform5}
\delta_{\xi}  A_{\mu\nu\lambda\rho\sigma} &=&\partial_{\mu}\xi_{\nu\lambda\rho\sigma}
-i g[A_{\mu},\xi_{\nu\lambda\rho\sigma}]
-i g \sum_{\nu \leftrightarrow \lambda\rho\sigma}
[A_{\mu\nu},\xi_{\lambda\rho\sigma}]-\nn\\
   &~&-ig\sum_{\nu\lambda \leftrightarrow \rho\sigma}
   [A_{\mu\nu\lambda},\xi_{\rho\sigma}]
      -ig\sum_{\nu\lambda\rho \leftrightarrow \sigma}
      [A_{\mu\nu\lambda\rho},\xi_{\sigma}]
-i g [A_{\mu\nu\lambda\rho},\xi], ~\nonumber
\eeqa
where the gauge parameters $\xi_{\nu\lambda\rho}$ and $\xi_{\nu\lambda\rho\sigma}$
are totally symmetric rank-3 and rank-4 tensors.
The extended gauge transformation of the higher-rank tensor gauge
fields induces the gauge transformation of the fields strengths of the form
(\ref{fieldstrenghparticulartransformation})
\beqa\label{spin4fieldstrenghthtransfor}
\delta G^{a}_{\mu\nu,\lambda\rho\sigma} =
g f^{abc} (~G^{b}_{\mu\nu,\lambda\rho\sigma} ~\xi^c  +
 G^{b}_{\mu\nu, \lambda\rho} ~\xi^{c}_{\sigma}+
G^{b}_{\mu\nu, \lambda\sigma} ~\xi^{c}_{\rho}+
G^{b}_{\mu\nu, \rho\sigma} ~\xi^{c}_{\lambda}+~~~~~~~~~~~~~~~~~~~~~~~~~\\
+ G^{b}_{\mu\nu,\lambda } ~\xi^{c}_{\rho\sigma}+
 G^{b}_{\mu\nu,\rho} ~\xi^{c}_{\lambda\sigma}+
 G^{b}_{\mu\nu,\sigma} ~\xi^{c}_{\lambda \rho}+
G^{b}_{\mu\nu } ~\xi^{c}_{\lambda\rho\sigma}~)\nonumber
\eeqa
and
\beqa\label{fieldstrengh5thtransfor}
\delta G^{a}_{\mu\nu,\lambda\rho\sigma\delta} =
g f^{abc} (~G^{b}_{\mu\nu,\lambda\rho\sigma\delta} ~\xi^c
&+& \sum_{ \lambda\rho,\sigma \leftrightarrow \delta}
G^{b}_{\mu\nu, \lambda\rho\sigma} ~\xi^{c}_{\delta}
+ \\
&+& \sum_{ \lambda\rho \leftrightarrow \sigma,\delta}
G^{b}_{\mu\nu, \lambda\rho} ~\xi^{c}_{\sigma\delta}+
         \sum_{ \lambda \leftrightarrow \rho,\sigma,\delta}
         G^{b}_{\mu\nu, \lambda} ~\xi^{c}_{\rho\sigma\delta}~+
         G^{b}_{\mu\nu } ~\xi^{c}_{\lambda\rho\sigma\delta}).\nonumber
\eeqa
Using the above homogeneous transformations for the field strengths
tensors one can demonstrate the invariance of the
Lagrangian ${{\cal L}}_3$ with respect to the extended gauge transformations
(see reference \cite{Savvidy:2005zm} for details).

The second invariant Lagrangian
can also be constructed explicitly in terms of the above field strength tensors.
The following seven Lorentz invariant quadratic forms
can be constructed by the corresponding field strength tensors
\cite{Savvidy:2005fi,Savvidy:2005ki,Savvidy:2005vm}
\beqa
G^{a}_{\mu\nu,\lambda\rho}G^{a}_{\mu\lambda,\nu\rho},~~~
G^{a}_{\mu\nu,\nu\lambda}G^{a}_{\mu\rho,\rho\lambda},~~~
G^{a}_{\mu\nu,\nu\lambda}G^{a}_{\mu\lambda,\rho\rho},~~~
G^{a}_{\mu\nu,\lambda}G^{a}_{\mu\lambda,\nu\rho\rho},~~~\nn\\
G^{a}_{\mu\nu,\lambda}G^{a}_{\mu\rho,\nu\lambda\rho},~~~
G^{a}_{\mu\nu,\nu}G^{a}_{\mu\lambda,\lambda\rho\rho},~~~
G^{a}_{\mu\nu}G^{a}_{\mu\lambda,\nu\lambda\rho\rho}.~~~~~~~~~~~~~~~~~
\eeqa
Calculating  the variation of each of these terms with respect to
the gauge transformation (\ref{fieldstrenghparticulartransformation}),
(\ref{spin4fieldstrenghthtransfor}) and  (\ref{fieldstrengh5thtransfor})
one can get convinced that the particular linear combination
\beqa\label{actionthreeprime}
{{\cal L}}^{'}_3 &=&  {1\over 4}
G^{a}_{\mu\nu,\lambda\rho}G^{a}_{\mu\lambda,\nu\rho}+
{1\over 4} G^{a}_{\mu\nu,\nu\lambda}G^{a}_{\mu\rho,\rho\lambda}+
{1\over 4}G^{a}_{\mu\nu,\nu\lambda}G^{a}_{\mu\lambda,\rho\rho}\nn\\
&+&{1\over 4}G^{a}_{\mu\nu,\lambda}G^{a}_{\mu\lambda,\nu\rho\rho}
+{1\over 2}G^{a}_{\mu\nu,\lambda}G^{a}_{\mu\rho,\nu\lambda\rho}
+{1\over 4}G^{a}_{\mu\nu,\nu}G^{a}_{\mu\lambda,\lambda\rho\rho}
+{1\over 4}G^{a}_{\mu\nu}G^{a}_{\mu\lambda,\nu\lambda\rho\rho}.
\eeqa
forms an invariant Lagrangian (see Appendix A).
In summary we have the following Lagrangian for the third-rank gauge field
$A^{a}_{\mu\nu\lambda}$
\beqa\label{actionthreeprimesum}
{{\cal L}}_3 + c_3 {{\cal L}}^{'}_3
=&-&{1\over 4}G^{a}_{\mu\nu,\lambda\rho}G^{a}_{\mu\nu,\lambda\rho}
-{1\over 8}G^{a}_{\mu\nu ,\lambda\lambda}G^{a}_{\mu\nu ,\rho\rho}
-{1\over 2}G^{a}_{\mu\nu,\lambda}  G^{a}_{\mu\nu ,\lambda \rho\rho}
-{1\over 8}G^{a}_{\mu\nu}  G^{a}_{\mu\nu ,\lambda \lambda\rho\rho}+ \nn\\
&+&c_3 \{ {1\over 4}
G^{a}_{\mu\nu,\lambda\rho}G^{a}_{\mu\lambda,\nu\rho}+
{1\over 4} G^{a}_{\mu\nu,\nu\lambda}G^{a}_{\mu\rho,\rho\lambda}+
{1\over 4}G^{a}_{\mu\nu,\nu\lambda}G^{a}_{\mu\lambda,\rho\rho}+\\
&+&{1\over 4}G^{a}_{\mu\nu,\lambda}G^{a}_{\mu\lambda,\nu\rho\rho}
+{1\over 2}G^{a}_{\mu\nu,\lambda}G^{a}_{\mu\rho,\nu\lambda\rho}
+{1\over 4}G^{a}_{\mu\nu,\nu}G^{a}_{\mu\lambda,\lambda\rho\rho}
+{1\over 4}G^{a}_{\mu\nu}G^{a}_{\mu\lambda,\nu\lambda\rho\rho} \},\nn
\eeqa
where $c_3$ is an arbitrary constant. Our intention is to investigate
the dependence of symmetries of the system
${{\cal L}}_3 + c_3 {{\cal L}}^{'}_3$ as a function of constant
$c_3$. The
system is always invariant with respect to the initial, extended gauge group
of transformations (\ref{fieldstrenghparticulartransformation}),
(\ref{spin4fieldstrenghthtransfor}) and  (\ref{fieldstrengh5thtransfor})
for any value of the constant $c_3$. We wish to know
if there exists a special value of the constant $c_3$ at which
the system will have even higher symmetry, as it happens in the case
of the rank-2 gauge field. We shall see that this indeed takes place.

\section{\it Enhanced Symmetry. Rank-3 Gauge Field}
The free part of the Lagrangian ${{\cal L}}_3$ comes from the terms
\beqa
-{1\over 4}G^{a}_{\mu\nu,\lambda\rho}G^{a}_{\mu\nu,\lambda\rho}
-{1\over 8}G^{a}_{\mu\nu ,\lambda\lambda}G^{a}_{\mu\nu ,\rho\rho}
\eeqa
and has the form
\beqa
{{\cal L}}^{free}_3=&-&{1 \over 2}\partial_{\mu}
A^{a}_{\nu \lambda\rho}\partial_{\mu} A^{a}_{\nu \lambda\rho}
+{1 \over 2}\partial_{\mu} A^{a}_{\nu \lambda\rho}\partial_{\nu}
A^{a}_{\mu \lambda\rho}
-{1 \over 4} \partial_{\mu} A^{a}_{\nu \lambda\lambda} \partial_{\mu }
A^{a}_{ \nu\rho\rho}
+{1 \over 4} \partial_{\mu} A^{a}_{\nu \lambda\lambda} \partial_{\nu}
A^{a}_{\mu \rho\rho}\nn\\
&=&{1 \over 2} A^{a}_{\alpha\alpha^{'}\alpha^{''}}
(\eta_{\alpha\gamma}\partial^{2} - \partial_{\alpha} \partial_{\gamma} )
({1 \over 2} \eta_{\alpha^{'} \gamma^{'}} \eta_{ \alpha^{''}  \gamma^{''} } +
{1 \over 2}\eta_{\alpha^{'}\gamma^{'}} \eta_{\alpha^{''} \gamma^{''}}+
{1 \over 2}\eta_{\alpha^{'}\alpha^{''}} \eta_{\gamma^{'} \gamma^{''}})
A^{a}_{\gamma\gamma^{'}\gamma^{''}} \nn\\
&=&{1 \over 2} A^{a}_{\alpha\alpha^{'}\alpha^{''}}
H_{\alpha\alpha^{'}\alpha^{''}\gamma\gamma^{'}\gamma^{''}}
A^{a}_{\gamma\gamma^{'}\gamma^{''}}  ,
\eeqa
where the quadratic form in the momentum representation is
$$
H_{\alpha\alpha^{'}\alpha^{''}\gamma\gamma^{'}\gamma^{''}}(k)=-{1 \over 2}
H_{\alpha\gamma}
(\eta_{\alpha^{'} \gamma^{'}} \eta_{ \alpha^{''}  \gamma^{''} } +
\eta_{\alpha^{'}\gamma^{''}} \eta_{\alpha^{''} \gamma^{'}}+
\eta_{\alpha^{'}\alpha^{''}} \eta_{\gamma^{'} \gamma^{''}}),
$$
where $H_{\alpha\gamma}= k^2 \eta_{\alpha\gamma} - k_{\alpha}k_{\gamma}$
and by construction is invariant with respect to the gauge
transformation
$$
\delta A^{a}_{\mu\nu\lambda} =\partial_{\mu} \xi^{a}_{\nu\lambda}
$$
because we have
\beqa
k_{\alpha}H_{\alpha\alpha^{'}\alpha^{''}\gamma\gamma^{'}\gamma^{''}}(k)=0.\nn
\eeqa
But it is not invariant with respect to the alternative gauge transformations
$$
\tilde{\delta} A^{a}_{\mu \nu\lambda} =\partial_{\nu} \zeta^{a}_{\mu\lambda}+
\partial_{\lambda} \zeta^{a}_{\mu\nu},
$$
where the gauge parameter $\zeta^{a}_{\mu\lambda}$ is a
totaly symmetric tensor.
This can be seen from the following relation in momentum representation
\beqa\label{firstlagrangiandivergence}
k_{\alpha^{'}}H_{\alpha\alpha^{'}\alpha^{''}\gamma\gamma^{'}\gamma^{''}}(k)&=&
-{1\over 2}H_{\alpha\gamma}~
(k_{\gamma^{'}} \eta_{ \alpha^{''}  \gamma^{''} } +
k_{\gamma^{''}} \eta_{\alpha^{''} \gamma^{'}}+
k_{\alpha^{''}} \eta_{\gamma^{'} \gamma^{''}}) \neq 0.
\eeqa
Let us consider now the free part of the Lagrangian ${{\cal L}}^{'}_3$ which comes from the terms
$$
 {1\over 4}
G^{a}_{\mu\nu,\lambda\rho}G^{a}_{\mu\lambda,\nu\rho}+
{1\over 4} G^{a}_{\mu\nu,\nu\lambda}G^{a}_{\mu\rho,\rho\lambda}+
{1\over 4}G^{a}_{\mu\nu,\nu\lambda}G^{a}_{\mu\lambda,\rho\rho},
$$
thus
\beqa\label{freeactionthreeprime}
{{\cal L}}^{'~free}_3=&+&{1 \over 4}\partial_{\mu}
A^{a}_{\nu \lambda\rho}\partial_{\mu} A^{a}_{\lambda\nu \rho}
-{1 \over 4}\partial_{\mu} A^{a}_{\nu \lambda\rho}\partial_{\lambda}
A^{a}_{\mu \nu\rho}
-{1 \over 4} \partial_{\nu} A^{a}_{\mu \lambda\rho} \partial_{\mu }
A^{a}_{ \lambda\nu\rho}
+{1 \over 4} \partial_{\nu} A^{a}_{ \mu\lambda\rho} \partial_{\lambda}
A^{a}_{\mu \nu\rho}\nn\\
&+&{1 \over 4}\partial_{\mu}
A^{a}_{\nu \nu\lambda }\partial_{\mu} A^{a}_{\rho\rho\lambda}
-{1 \over 4}\partial_{\mu} A^{a}_{\nu \nu\lambda}\partial_{\rho}
A^{a}_{\mu \rho\lambda}
-{1 \over 4} \partial_{\nu} A^{a}_{\mu \nu\lambda } \partial_{\mu }
A^{a}_{\rho \rho\lambda }
+{1 \over 4} \partial_{\nu} A^{a}_{ \mu\nu\lambda} \partial_{\rho}
A^{a}_{\mu \rho\lambda}\nn\\
&+&{1 \over 4}\partial_{\mu}
A^{a}_{\nu \nu\lambda }\partial_{\mu} A^{a}_{\lambda\rho\rho}
-{1 \over 4}\partial_{\mu} A^{a}_{\nu \nu\lambda}\partial_{\lambda}
A^{a}_{\mu \rho\rho}
-{1 \over 4} \partial_{\nu} A^{a}_{\mu \nu\lambda } \partial_{\mu }
A^{a}_{\lambda \rho \rho}
+{1 \over 4} \partial_{\nu} A^{a}_{ \mu\nu\lambda} \partial_{\lambda}
A^{a}_{\mu \rho\rho}\nn\\
&=&{1 \over 2} A^{a}_{\alpha\alpha^{'}\alpha^{''}}
H^{~'}_{\alpha\alpha^{'}\alpha^{''}\gamma\gamma^{'}\gamma^{''}}
A^{a}_{\gamma\gamma^{'}\gamma^{''}}  ,
\eeqa
where one should symmetrize the
$H^{~'}_{\alpha\alpha^{'}\alpha^{''}\gamma\gamma^{'}\gamma^{''}}$  over the
$\alpha^{'} \leftrightarrow\alpha^{''}$,
$\gamma^{'} \leftrightarrow \gamma^{''}$ and the exchange of two sets
of indices
$\alpha \alpha^{'}\alpha^{''} \leftrightarrow \gamma\gamma^{'}\gamma^{''}$,
so that the second quadratic form in the momentum representation is
(see also Appendix B for derivation)
\beqa
H^{~'}_{\alpha\alpha^{'}\alpha^{''}\gamma\gamma^{'}\gamma^{''}}(k)=
{1\over 8}\{ &+& (k^2 \eta_{\alpha\alpha^{'}}-k_{\alpha}k_{\alpha^{'}})
(\eta_{\alpha^{''}\gamma} \eta_{\gamma^{'} \gamma^{''}}
+\eta_{\alpha^{''}\gamma^{'}} \eta_{\gamma \gamma^{''}}
+\eta_{\alpha^{''}\gamma^{''}} \eta_{\gamma\gamma^{'} })\nn\\
&+& (k^2 \eta_{\alpha\alpha^{''}} -k_{\alpha}k_{\alpha^{''}})
(\eta_{\alpha^{'}\gamma} \eta_{\gamma^{'} \gamma^{''}}
+\eta_{\alpha^{'}\gamma^{'}} \eta_{\gamma \gamma^{''}}
+\eta_{\alpha^{'}\gamma^{''}} \eta_{\gamma\gamma^{'} })\nn\\
&+&(k^2 \eta_{\alpha\gamma^{'}}-k_{\alpha}k_{\gamma^{'}})
(\eta_{\alpha^{'}\gamma} \eta_{\alpha^{''} \gamma^{''}}
+\eta_{\alpha^{'}\gamma^{''}} \eta_{\alpha^{''}\gamma }
+\eta_{\alpha^{'}\alpha^{''}} \eta_{\gamma\gamma^{''} })\nn\\
&+&(k^2 \eta_{\alpha\gamma^{''}}-k_{\alpha}k_{\gamma^{''}})
(\eta_{\alpha^{'}\gamma} \eta_{\alpha^{''} \gamma^{'}}
+\eta_{\alpha^{'}\gamma^{'}} \eta_{\alpha^{''}\gamma }+
\eta_{\alpha^{'}\alpha^{''}} \eta_{\gamma \gamma^{'}})~\}\nn\\
             -{ 1\over 8}\{&+&k_{\gamma}k_{\alpha^{'}}
(\eta_{\alpha\gamma^{'}} \eta_{\alpha^{''} \gamma^{''}}
+\eta_{\alpha\gamma^{''}} \eta_{\alpha^{''} \gamma^{'}}
+\eta_{\alpha\alpha^{''}} \eta_{\gamma^{'}\gamma^{''} })\nn\\
&+&k_{\gamma}k_{\alpha^{''}}
(\eta_{\alpha\gamma^{'}} \eta_{\alpha^{'} \gamma^{''}}
+\eta_{\alpha\gamma^{''}} \eta_{\alpha^{'}\gamma^{'} }
+\eta_{\alpha\alpha^{'}} \eta_{\gamma^{'}\gamma^{''} })\nn\\
&+&k_{\gamma}k_{\gamma^{'}}
(\eta_{\alpha\alpha^{'}} \eta_{\alpha^{''} \gamma^{''}}
+\eta_{\alpha\alpha^{''}} \eta_{\alpha^{'} \gamma^{''}}
+\eta_{\alpha\gamma^{''}} \eta_{\alpha^{'}\alpha^{''} })\nn\\
&+&k_{\gamma}k_{\gamma^{''}}
(\eta_{\alpha\alpha^{'}} \eta_{\alpha^{''} \gamma^{'}}
+\eta_{\alpha\alpha^{''}} \eta_{\alpha^{'}\gamma^{'} }
+\eta_{\alpha\gamma^{'}} \eta_{\alpha^{'}\alpha^{''} })~\}\nn\\
   +{ 1\over 4}\{&+&\eta_{\alpha\gamma} (k_{\alpha^{'}}k_{\gamma^{'}}
\eta_{\alpha^{''} \gamma^{''}} + k_{\alpha^{'}}k_{\gamma^{''}}
\eta_{\alpha^{''} \gamma^{'}} + k_{\alpha^{''}}k_{\gamma^{'}}
\eta_{\alpha^{'} \gamma^{''}} \nn\\
&+&k_{\alpha^{''}}k_{\gamma^{''}}
\eta_{\alpha^{'} \gamma^{'}} + k_{\alpha^{'}}k_{\alpha^{''}}
\eta_{ \gamma^{'}\gamma^{''}} +k_{\gamma^{'}}k_{\gamma^{''}}
\eta_{\alpha^{'}\alpha^{''} })~\}.
\eeqa
It is again invariant with respect  to the transformation $\delta A^{a}_{\mu\nu\lambda} =
\partial_{\mu} \xi^{a}_{\nu\lambda}$ because we have
\beqa
k_{\alpha}H^{~'}_{\alpha\alpha^{'}\alpha^{''}\gamma\gamma^{'}\gamma^{''}}(k)&=&0,\nn
\eeqa
but it is not invariant with respect to the transformation
$\tilde{\delta} A^{a}_{\mu \nu\lambda} =\partial_{\nu} \zeta^{a}_{\mu\lambda}
+\partial_{\lambda} \zeta^{a}_{\mu\nu}$,
as one can see from the analogous relation (see also Appendix B for derivation)
\beqa\label{secondlagrangiandivergence}
k_{\alpha^{'}}H^{~'}_{\alpha\alpha^{'}\alpha^{''}\gamma\gamma^{'}\gamma^{''}}(k)=
{1\over 8}\{&+&H_{\alpha\alpha^{''}}
(k_{\gamma^{'}} \eta_{\gamma \gamma^{''}}
+k_{\gamma^{''}} \eta_{\gamma\gamma^{'} })\nn\\
&+&H_{\alpha\gamma^{'}}
(k_{\gamma^{''}} \eta_{\alpha^{''}\gamma }
+k_{\alpha^{''}} \eta_{\gamma\gamma^{''} })\nn\\
&+&H_{\alpha\gamma^{''}}
(k_{\gamma^{'}} \eta_{\alpha^{''}\gamma }+
k_{\alpha^{''}} \eta_{\gamma \gamma^{'}})~\}\nn\\
             -{ 1\over 4}\{
&+&k_{\gamma}k_{\alpha^{''}}
(k_{\gamma^{''}}\eta_{\alpha\gamma^{'}}
+k_{\gamma^{'} }\eta_{\alpha\gamma^{''}})
+k_{\gamma}k_{\gamma^{'}}
k_{\gamma^{''}}\eta_{\alpha\alpha^{''}}
-3\eta_{\alpha\gamma}k_{\alpha^{''}}k_{\gamma^{'}}k_{\gamma^{''}} ~\}\nn\\
   +{ 1\over 4}\{&+&H_{\alpha\gamma}
   (k_{\gamma^{'}}\eta_{\alpha^{''} \gamma^{''}}+
   k_{\gamma^{''}}\eta_{\alpha^{''} \gamma^{'}}+ k_{\alpha^{''}}
\eta_{ \gamma^{'}\gamma^{''}})~\} \neq 0.
\eeqa
We have to see now whether the sum
$$
k_{\alpha^{'}} (~H^{~'}_{\alpha\alpha^{'}\alpha^{''}\gamma\gamma^{'}\gamma^{''}}+
c_3 H^{~'}_{\alpha\alpha^{'}\alpha^{''}\gamma\gamma^{'}\gamma^{''}}~)
$$
can be made equal to zero by an appropriate choice of the coefficient $c_3$.
For that let us compare the expressions (\ref{firstlagrangiandivergence}) and
(\ref{secondlagrangiandivergence}) for divergences. As one can see, only the last term in
(\ref{secondlagrangiandivergence})
$$
H_{\alpha\gamma}
   (k_{\gamma^{'}}\eta_{\alpha^{''} \gamma^{''}}+
   k_{\gamma^{''}}\eta_{\alpha^{''} \gamma^{'}}+ k_{\alpha^{''}}
\eta_{ \gamma^{'}\gamma^{''}})
$$
and the whole term (\ref{firstlagrangiandivergence}) can  cancel each other if
we choose $c_3=2$, but this will leave the rest of the terms in (\ref{secondlagrangiandivergence})
untouched, thus
$$
k_{\alpha^{'}} (~H^{~'}_{\alpha\alpha^{'}\alpha^{''}\gamma\gamma^{'}\gamma^{''}}+
c_3 H^{~'}_{\alpha\alpha^{'}\alpha^{''}\gamma\gamma^{'}\gamma^{''}}~) \neq 0
$$
for all values of $c_3$. This situation differs from the case of
the rank-2 gauge field $A^{a}_{\mu\nu}$. In the last case we were able
to choose the coefficient $c_2 =1$ so that the divergences (\ref{currentdivergence}) and
(\ref{currentdivergenceprime}) cancel each other and we got (\ref{zeroderivatives})
$$
k_{\acute{\alpha}}(H_{\alpha\acute{\alpha}\gamma\acute{\gamma}}+
H^{'}_{\alpha\acute{\alpha}\gamma\acute{\gamma}})=0.
$$
Therefore the Lagrangian $\CL^{free}_2 + \CL^{free~'}_2 $ has enhanced
invariance with respect to a  large gauge group of transformations
$\delta A^{a}_{\mu \lambda} =\partial_{\mu} \xi^{a}_{\lambda}+
\partial_{\lambda} \eta^{a}_{\mu}$.

In order to understand the reason why in the case of the rank-3 gauge field
it is impossible fully cancel divergences
we have to analyze  the corresponding field equations.
We shall compare the resulting equation with the
equation derived by Schwinger \cite{schwinger}
for the totally symmetric Abelian rank-3 tensor field
in order to get better insight into the problem.
It has been proved by Schwinger \cite{schwinger} that it is impossible to derive
field equation for the totally symmetric rank-3 tensor which is invariant
with respect to the full gauge group of  transformations
$\delta A_{\mu \nu\lambda} =\partial_{\mu} \xi^{a}_{\nu\lambda}+
\partial_{\nu} \xi_{\mu\lambda}+ \partial_{\lambda} \xi_{\mu\nu}$
without imposing some restriction on the gauge parameters $\xi_{\mu\nu}$.
As he demonstrated, the gauge parameter should be traceless $\xi_{\mu\mu}=0$.
We shall see that similar phenomena take place also in our case, that is,
the second gauge parameter $\zeta^{a}_{\mu\lambda}$ should
fulfill constraint which we shall derive below (see equation
(\ref{restrictionongaugeparameters})).

What we would like to prove is
that the equation has enhanced invariance
with respect to the gauge group of transformations
\be\label{enhancedgaugetransformationrank3}
\tilde{\delta} A^{a}_{\mu \nu\lambda} =\partial_{\nu} \zeta^{a}_{\mu\lambda}+
\partial_{\lambda} \zeta^{a}_{\mu\nu},
\ee
but in our case the gauge parameters $\zeta^{a}_{\mu\lambda}$ should fulfill the following
constraint:
\be\label{restrictionongaugeparameters}
\partial_{\rho}\zeta^{a}_{\rho\lambda}-\partial_{\lambda} \zeta^{a}_{ \rho\rho}=0.
\ee
This takes place when we choose the coefficient
$
c_3 = 4/3~.
$
Indeed, let us consider the equation of motion.
From the first form ${{\cal L}}^{free}_3$ we have the following
contribution to the field equation:
\beqa
H_{\alpha\alpha^{'}\alpha^{''}\gamma\gamma^{'}\gamma^{''}}
A^{a}_{\gamma\gamma^{'}\gamma^{''}}
=\partial^{2} A^{a}_{\alpha\alpha^{'}\alpha^{''}} -\partial_{\alpha} \partial_{\rho}
A^{a}_{\rho\alpha^{'}\alpha^{''}} +
{1 \over 2}\eta_{\alpha^{'}\alpha^{''}}( \partial^{2} A^{a}_{\alpha\rho\rho}
-\partial_{\alpha} \partial_{\rho} A^{a}_{\rho\lambda\lambda}),
\eeqa
and from the second one ${{\cal L}}^{'~free}_3$ we have
\beqa
H^{'}_{\alpha\alpha^{'}\alpha^{''}\gamma\gamma^{'}\gamma^{''}}
A^{a}_{\gamma\gamma^{'}\gamma^{''}}=-{1 \over 8}\{ \partial^{2}( A^{a}_{\alpha^{'}\alpha \alpha^{''}}+
A^{a}_{\alpha^{'}\alpha^{''}\alpha }+A^{a}_{\alpha^{''}\alpha \alpha^{'}}+
A^{a}_{\alpha^{''}\alpha^{'}\alpha })-\nn\\
-\partial_{\alpha} \partial_{\rho}(A^{a}_{\alpha^{'}\rho\alpha^{''}}
+A^{a}_{\alpha^{'}\alpha^{''}\rho}
+A^{a}_{\alpha^{''}\rho\alpha^{'}}
+A^{a}_{\alpha^{''}\alpha^{'}\rho})-\nn\\
-\partial_{\alpha^{'}} \partial_{\rho}(A^{a}_{\rho\alpha\alpha^{''}}
+A^{a}_{\rho\alpha^{''}\alpha} - A^{a}_{\alpha\rho\alpha^{''}}
-A^{a}_{\alpha\alpha^{''}\rho}-A^{a}_{\alpha\alpha^{''}\rho}
-A^{a}_{\alpha\rho\alpha^{''}})-\nn\\
-\partial_{\alpha^{''}} \partial_{\rho}(A^{a}_{\rho\alpha\alpha^{'}}
+A^{a}_{\rho\alpha^{'}\alpha} - A^{a}_{\alpha\rho\alpha^{'}}
-A^{a}_{\alpha\alpha^{'}\rho}-A^{a}_{\alpha\alpha^{'}\rho}
-A^{a}_{\alpha\rho\alpha^{'}})-\nn\\
-\partial_{\alpha}\partial_{\alpha^{'}} (A^{a}_{\alpha^{''}\rho\rho}
+A^{a}_{\rho\alpha^{''}\rho} + A^{a}_{\rho\rho\alpha^{''}})
-\partial_{\alpha}\partial_{\alpha^{''}} (A^{a}_{\alpha^{'}\rho\rho}
+A^{a}_{\rho\alpha^{'}\rho} + A^{a}_{\rho\rho\alpha^{'}})
+2 \partial_{\alpha^{'}}\partial_{\alpha^{''}} A^{a}_{\alpha\rho\rho}\}-\nn\\
    -{1 \over 8}\{\eta_{\alpha\alpha^{'}} [\partial^{2}( A^{a}_{\alpha^{''}\rho\rho}+
A^{a}_{\rho\alpha^{''}\rho}+A^{a}_{\rho\rho\alpha^{''}})-
\partial_{\alpha^{''}}\partial_{\rho} A^{a}_{\rho\lambda \lambda}-
\partial_{\lambda}\partial_{\rho} (A^{a}_{\rho \alpha^{''}\lambda}
+ A^{a}_{\rho \lambda\alpha^{''}})]+\nn\\
+\eta_{\alpha\alpha^{''}} [\partial^{2}( A^{a}_{\alpha^{'}\rho\rho}+
A^{a}_{\rho\alpha^{'}\rho}+A^{a}_{\rho\rho\alpha^{'}})-
\partial_{\alpha^{'}}\partial_{\rho} A^{a}_{\rho\lambda \lambda}-
\partial_{\lambda}\partial_{\rho} (A^{a}_{\rho \alpha^{'}\lambda}
+ A^{a}_{\rho \lambda\alpha^{'}})]+~~~\nn\\
+\eta_{\alpha^{'}\alpha^{''}}[ \partial^{2}( A^{a}_{\rho\alpha\rho}+
A^{a}_{\rho\rho\alpha})
-\partial_{\alpha} \partial_{\rho}( A^{a}_{\lambda\rho\lambda} +A^{a}_{\lambda\lambda} \rho)
-\partial_{\lambda} \partial_{\rho}( A^{a}_{\rho\alpha\lambda} + A^{a}_{\lambda\rho\alpha}
-2A^{a}_{\alpha\lambda\rho})] \}.
\eeqa
Summing these two pieces together we shall get the following free field equation of
motion for the rank-3 tensor gauge field:
\beqa
(H_{\alpha\alpha^{'}\alpha^{''}\gamma\gamma^{'}\gamma^{''}}
+c_3 H^{~'}_{\alpha\alpha^{'}\alpha^{''}\gamma\gamma^{'}\gamma^{''}})
A^{a}_{\gamma\gamma^{'}\gamma^{''}}=
\partial^{2}( A^{a}_{\alpha \alpha^{'}\alpha^{''}}- {c_3\over 4}
A^{a}_{\alpha^{'}\alpha^{''}\alpha }
- {c_3\over 4} A^{a}_{\alpha^{''}\alpha \alpha^{'}})-\nn\\
-\partial_{\alpha} \partial_{\rho}(A^{a}_{\rho\alpha^{'}\alpha^{''}}
- {c_3\over 4}A^{a}_{\alpha^{'}\alpha^{''}\rho }
- {c_3\over 4} A^{a}_{\alpha^{''}\rho \alpha^{'}})
-{c_3\over 4}\partial_{\alpha^{'}} \partial_{\rho}(A^{a}_{\alpha\rho\alpha^{''}}
+A^{a}_{\alpha\alpha^{''}\rho} - A^{a}_{\rho\alpha\alpha^{''}})-\nn\\
-{c_3\over 4}\partial_{\alpha^{''}} \partial_{\rho}(A^{a}_{\alpha\rho\alpha^{'}}
+A^{a}_{\alpha\alpha^{'}\rho} - A^{a}_{\rho\alpha\alpha^{'}})
+{c_3\over 8}\partial_{\alpha}\partial_{\alpha^{'}} (A^{a}_{\alpha^{''}\rho\rho}
+A^{a}_{\rho\alpha^{''}\rho} + A^{a}_{\rho\rho\alpha^{''}})+\nn\\
+{c_3\over 8}\partial_{\alpha}\partial_{\alpha^{''}} (A^{a}_{\alpha^{'}\rho\rho}
+A^{a}_{\rho\alpha^{'}\rho} + A^{a}_{\rho\rho\alpha^{'}})
-{c_3\over 4}\partial_{\alpha^{'}}\partial_{\alpha^{''}} A^{a}_{\alpha\rho\rho}-\nn\\
-{c_3 \over 8}\eta_{\alpha\alpha^{'}} (\partial^{2} A^{a}_{\alpha^{''}\rho\rho}-
\partial_{\alpha^{''}}\partial_{\rho} A^{a}_{\rho\lambda \lambda}
+2 \partial^{2}A^{a}_{\rho\rho\alpha^{''}}-2
\partial_{\lambda}\partial_{\rho} A^{a}_{\rho \lambda\alpha^{''}})-\nn\\
-{c_3 \over 8}\eta_{\alpha\alpha^{''}} (\partial^{2} A^{a}_{\alpha^{'}\rho\rho}-
\partial_{\alpha^{'}}\partial_{\rho} A^{a}_{\rho\lambda \lambda}
+2 \partial^{2}A^{a}_{\rho\rho\alpha^{'}}-2
\partial_{\lambda}\partial_{\rho} A^{a}_{\rho \lambda\alpha^{'}})-\nn\\
+{1 \over 2}\eta_{\alpha^{'}\alpha^{''}}( \partial^{2} A^{a}_{\alpha\rho\rho}
-\partial_{\alpha} \partial_{\rho} A^{a}_{\rho\lambda\lambda}
-{c_3 \over 2}\partial^{2} A^{a}_{\rho\rho\alpha}
+{c_3 \over 2}\partial_{\alpha} \partial_{\rho} A^{a}_{\lambda\lambda\rho}
-{c_3 \over 2}\partial_{\lambda} \partial_{\rho} A^{a}_{\alpha\lambda\rho}
+ {c_3 \over 2}\partial_{\lambda} \partial_{\rho} A^{a}_{\lambda\rho\alpha})
=0. \nn
\eeqa
We shall prove that this equation is invariant with respect to the
gauge transformation
$$
\tilde{\delta} A^{a}_{\mu \nu\lambda} =\partial_{\nu} \zeta^{a}_{\mu\lambda}+
\partial_{\lambda} \zeta^{a}_{\mu\nu}
$$
if we choose the coefficient $c_3 = 4/3$.
Performing the above gauge transformation of the field
one can see that the terms which originate from
differential operators $\partial^{2},~$
$\partial_{\alpha} \partial_{\rho},~~$$\partial_{\alpha^{'}} \partial_{\rho}$
and $\partial_{\alpha^{''}} \partial_{\rho}$ in the above equation
cancel each other if we choose
\be
c_3={4\over 3}~~.
\ee
The rest of the terms have the following form:
\beqa
(H_{\alpha\alpha^{'}\alpha^{''}\gamma\gamma^{'}\gamma^{''}}
+{4\over 3} H^{~'}_{\alpha\alpha^{'}\alpha^{''}\gamma\gamma^{'}\gamma^{''}})
\tilde{\delta} A^{a}_{\gamma\gamma^{'}\gamma^{''}}=\nn\\
+{1\over 3}\partial_{\alpha}\partial_{\alpha^{'}} \partial_{\rho}  \zeta^{a}_{\rho\alpha^{''}}
+{1\over 3}\partial_{\alpha}\partial_{\alpha^{''}} \partial_{\rho}  \zeta^{a}_{\rho\alpha^{'}}
-{4\over 3}\partial_{\alpha^{'}}\partial_{\alpha^{''}}\partial_{\rho} \zeta^{a}_{\rho\alpha}
+{2\over 3}\partial_{\alpha}\partial_{\alpha^{'}}\partial_{\alpha^{''}}  \zeta^{a}_{\rho\rho}\nn\\
-{1 \over 6}\eta_{\alpha\alpha^{'}} (2 \partial_{\rho} \partial^{2} \zeta^{a}_{\rho\alpha^{''}}
-4\partial_{\alpha^{''}}\partial_{\lambda} \partial_{\rho} \zeta^{a}_{\lambda \rho}
+2 \partial_{\alpha^{''}}\partial^{2} \zeta^{a}_{\rho\rho})\nn\\
-{1 \over 6}\eta_{\alpha\alpha^{''}} (2 \partial_{\rho} \partial^{2} \zeta^{a}_{\rho\alpha^{'}}
-4\partial_{\alpha^{'}}\partial_{\lambda} \partial_{\rho} \zeta^{a}_{\lambda \rho}
+2 \partial_{\alpha^{'}}\partial^{2} \zeta^{a}_{\rho\rho})\nn\\
+{1 \over 3}\eta_{\alpha^{'}\alpha^{''}}(\partial_{\rho} \partial^{2} \zeta^{a}_{\rho\alpha}
-\partial_{\alpha} \partial_{\lambda}\partial_{\rho} \zeta^{a}_{\lambda\rho})
\eeqa
and can be rewritten in the form which makes  the desired invariance explicit:
\beqa
(H_{\alpha\alpha^{'}\alpha^{''}\gamma\gamma^{'}\gamma^{''}}
+{4\over 3} H^{~'}_{\alpha\alpha^{'}\alpha^{''}\gamma\gamma^{'}\gamma^{''}})
\delta A^{a}_{\gamma\gamma^{'}\gamma^{''}}=
+{1\over 3}\partial_{\alpha}\partial_{\alpha^{'}}( \partial_{\rho}  \zeta^{a}_{\rho\alpha^{''}}
-\partial_{\alpha^{''}}  \zeta^{a}_{\rho\rho})\nn\\
+{1\over 3}\partial_{\alpha}\partial_{\alpha^{''}} (\partial_{\rho}  \zeta^{a}_{\rho\alpha^{'}}
-\partial_{\alpha^{'}}  \zeta^{a}_{\rho\rho})
-{4\over 3}\partial_{\alpha^{'}}\partial_{\alpha^{''}}(\partial_{\rho} \zeta^{a}_{\rho\alpha}
-\partial_{\alpha}  \zeta^{a}_{\rho\rho})\nn\\
-{1 \over 3}\eta_{\alpha\alpha^{'}} [ \partial^{2}( \partial_{\rho}  \zeta^{a}_{\rho\alpha^{''}}
-\partial_{\alpha^{''}}\zeta^{a}_{\rho\rho})+2
\partial_{\alpha^{''}}\partial_{\lambda} (\partial_{\lambda} \zeta^{a}_{ \rho\rho}-
\partial_{\rho}\zeta^{a}_{\rho\lambda})]\nn\\
-{1 \over 3}\eta_{\alpha\alpha^{''}} [ \partial^{2}( \partial_{\rho}  \zeta^{a}_{\rho\alpha^{'}}
-\partial_{\alpha^{'}}\zeta^{a}_{\rho\rho})+2
\partial_{\alpha^{'}}\partial_{\lambda} (\partial_{\lambda} \zeta^{a}_{ \rho\rho}-
\partial_{\rho}\zeta^{a}_{\rho\lambda})]\nn\\
+{1 \over 3}\eta_{\alpha^{'}\alpha^{''}} [ \partial^{2}(\partial_{ \rho}\zeta^{a}_{\rho\alpha}
-\partial_{\alpha}  \zeta^{a}_{\rho\rho})+
\partial_{\alpha}\partial_{\lambda} (\partial_{\lambda} \zeta^{a}_{ \rho\rho}-
\partial_{\rho}\zeta^{a}_{\rho\lambda})].
\eeqa
From that we see that if the gauge parameter satisfies the conditions (\ref{restrictionongaugeparameters})
$$
\partial_{\rho}\zeta^{a}_{\rho\lambda}-\partial_{\lambda} \zeta^{a}_{ \rho\rho}=0
$$
the equation is indeed invariant with respect to a larger group of gauge
transformations
$
\tilde{\delta} A^{a}_{\mu \nu\lambda} =\partial_{\nu} \zeta^{a}_{\mu\lambda}+
\partial_{\lambda} \zeta^{a}_{\mu\nu},
$
because
\be
(H_{\alpha\alpha^{'}\alpha^{''}\gamma\gamma^{'}\gamma^{''}}
+{4\over 3} H^{~'}_{\alpha\alpha^{'}\alpha^{''}\gamma\gamma^{'}\gamma^{''}})
\tilde{\delta} A^{a}_{\gamma\gamma^{'}\gamma^{''}}=0.
\ee
The final form of the equation is
\beqa\label{freethirdrankequations}
\partial^{2}( A^{a}_{\alpha \alpha^{'}\alpha^{''}}
-{1\over 3}A^{a}_{\alpha^{'}\alpha^{''}\alpha }
- {1\over 3} A^{a}_{\alpha^{''}\alpha \alpha^{'}})
-\partial_{\alpha} \partial_{\rho}(A^{a}_{\rho\alpha^{'}\alpha^{''}}
- {1\over 3}A^{a}_{\alpha^{'}\alpha^{''}\rho }
- {1\over 3} A^{a}_{\alpha^{''}\rho \alpha^{'}})-\\
-{1\over 3}\partial_{\alpha^{'}} \partial_{\rho}(A^{a}_{\alpha\rho\alpha^{''}}
+A^{a}_{\alpha\alpha^{''}\rho} - A^{a}_{\rho\alpha\alpha^{''}})
-{1\over 3}\partial_{\alpha^{''}} \partial_{\rho}(A^{a}_{\alpha\rho\alpha^{'}}
+A^{a}_{\alpha\alpha^{'}\rho} - A^{a}_{\rho\alpha\alpha^{'}})+\nn\\
+{1\over 6}\partial_{\alpha}\partial_{\alpha^{'}} (A^{a}_{\alpha^{''}\rho\rho}
+A^{a}_{\rho\alpha^{''}\rho} + A^{a}_{\rho\rho\alpha^{''}})
+{1\over 6}\partial_{\alpha}\partial_{\alpha^{''}} (A^{a}_{\alpha^{'}\rho\rho}
+A^{a}_{\rho\alpha^{'}\rho} + A^{a}_{\rho\rho\alpha^{'}})
-{1\over 3}\partial_{\alpha^{'}}\partial_{\alpha^{''}} A^{a}_{\alpha\rho\rho}-\nn\\
-{1 \over 6}\eta_{\alpha\alpha^{'}} (\partial^{2} A^{a}_{\alpha^{''}\rho\rho}-
\partial_{\alpha^{''}}\partial_{\rho} A^{a}_{\rho\lambda \lambda}
+2 \partial^{2}A^{a}_{\rho\rho\alpha^{''}}-2
\partial_{\lambda}\partial_{\rho} A^{a}_{\rho \lambda\alpha^{''}})-\nn\\
-{1 \over 6}\eta_{\alpha\alpha^{''}} (\partial^{2} A^{a}_{\alpha^{'}\rho\rho}-
\partial_{\alpha^{'}}\partial_{\rho} A^{a}_{\rho\lambda \lambda}
+2 \partial^{2}A^{a}_{\rho\rho\alpha^{'}}-2
\partial_{\lambda}\partial_{\rho} A^{a}_{\rho \lambda\alpha^{'}})-\nn\\
+{1 \over 2}\eta_{\alpha^{'}\alpha^{''}}( \partial^{2} A^{a}_{\alpha\rho\rho}
-\partial_{\alpha} \partial_{\rho} A^{a}_{\rho\lambda\lambda}
-{2 \over 3}\partial^{2} A^{a}_{\rho\rho\alpha}
+{2 \over 3}\partial_{\alpha} \partial_{\rho} A^{a}_{\lambda\lambda\rho}
-{2 \over 3}\partial_{\lambda} \partial_{\rho} A^{a}_{\alpha\lambda\rho}
+ {2 \over 3}\partial_{\lambda} \partial_{\rho} A^{a}_{\lambda\rho\alpha})
=0\nn
\eeqa
and it is invariant with respect to the gauge group of transformations
\be\label{fullgroupofextendedtransformation}
\delta A^{a}_{\mu\nu\lambda} =\partial_{\mu} \xi^{a}_{\nu\lambda},~~~~~~~~~
\tilde{\delta} A^{a}_{\mu \nu\lambda} =\partial_{\nu} \zeta^{a}_{\mu\lambda}+
\partial_{\lambda} \zeta^{a}_{\mu\nu},~~~~~~~~
\partial_{\rho}\zeta^{a}_{\rho\lambda}-\partial_{\lambda} \zeta^{a}_{ \rho\rho}=0.
\ee
One should stress that there are no restrictions on the gauge parameters $\xi^{a}_{\mu\nu}$.
The above invariance of the equation now can be checked directly without
referring to the previous analysis.
In summary, we have the following Lagrangian for the third-rank gauge field
$A^{a}_{\mu\nu\lambda}$:
\beqa\label{actionthreeprimesum}
{{\cal L}}_3 + {4 \over 3} {{\cal L}}^{'}_3
=&-&{1\over 4}G^{a}_{\mu\nu,\lambda\rho}G^{a}_{\mu\nu,\lambda\rho}
-{1\over 8}G^{a}_{\mu\nu ,\lambda\lambda}G^{a}_{\mu\nu ,\rho\rho}
-{1\over 2}G^{a}_{\mu\nu,\lambda}  G^{a}_{\mu\nu ,\lambda \rho\rho}
-{1\over 8}G^{a}_{\mu\nu}  G^{a}_{\mu\nu ,\lambda \lambda\rho\rho}+ \nn\\
&+&{1\over 3}G^{a}_{\mu\nu,\lambda\rho}G^{a}_{\mu\lambda,\nu\rho}+
{1\over 3} G^{a}_{\mu\nu,\nu\lambda}G^{a}_{\mu\rho,\rho\lambda}+
{1\over 3}G^{a}_{\mu\nu,\nu\lambda}G^{a}_{\mu\lambda,\rho\rho}+\\
&+&{1\over 3}G^{a}_{\mu\nu,\lambda}G^{a}_{\mu\lambda,\nu\rho\rho}
+{2\over 3}G^{a}_{\mu\nu,\lambda}G^{a}_{\mu\rho,\nu\lambda\rho}
+{1\over 3}G^{a}_{\mu\nu,\nu}G^{a}_{\mu\lambda,\lambda\rho\rho}
+{1\over 3}G^{a}_{\mu\nu}G^{a}_{\mu\lambda,\nu\lambda\rho\rho}.\nn
\eeqa
We shall present the free equation of motion (\ref{freethirdrankequations})
also in terms of field strength
tensors. The kinetic term of the above Lagrangian is
\beqa\label{freeactionthreeprimesum}
{{\cal L}}_3 + {4 \over 3} {{\cal L}}^{'}_3 ~\vert_{free}
=&-&{1\over 4}F^{a}_{\mu\nu,\lambda\rho}F^{a}_{\mu\nu,\lambda\rho}
-{1\over 8}F^{a}_{\mu\nu ,\lambda\lambda}F^{a}_{\mu\nu ,\rho\rho}\nn\\
&+&{1\over 3}F^{a}_{\mu\nu,\lambda\rho}F^{a}_{\mu\lambda,\nu\rho}+
{1\over 3} F^{a}_{\mu\nu,\nu\lambda}F^{a}_{\mu\rho,\rho\lambda}+
{1\over 3}F^{a}_{\mu\nu,\nu\lambda}F^{a}_{\mu\lambda,\rho\rho},
\eeqa
where
$$
F^{a}_{\mu\nu ,\lambda \rho} =
\partial_{\mu} A^{a}_{\nu \lambda \rho} - \partial_{\nu} A^{a}_{\mu \lambda\rho}.
$$
The variation of the above  Lagrangian over the field
$A^{a}_{\nu \lambda \rho}$ gives
the free equation   written in terms
of field strength tensor $F^{a}_{\mu\nu ,\lambda \rho}$ and
it is identical to the equation (\ref{freethirdrankequations})
\beqa\label{freeequation}
\partial_{\mu}F^{a}_{\mu\nu,\lambda\rho}
-{1\over 3}\partial_{\mu}F^{a}_{\mu\lambda,\nu\rho}
- {1\over 3} \partial_{\mu}F^{a}_{\mu\rho,\nu\lambda}
+ {1\over 3}\partial_{\mu}F^{a}_{\nu\lambda,\mu\rho}
+{1\over 3} \partial_{\mu}F^{a}_{\nu\rho,\mu\lambda} +\nn\\
+{1\over 3}\partial_{\lambda}F^{a}_{\nu\mu,\mu\rho}
+{1\over 3}\partial_{\rho}F^{a}_{\nu\mu,\mu\lambda}
+{1\over 6}\partial_{\lambda}F^{a}_{\nu\rho,\mu\mu}
+{1\over 6}\partial_{\rho}F^{a}_{\nu\lambda,\mu\mu}-\nn\\
-\eta_{\lambda\nu} ({1\over 3}\partial_{\mu}F^{a}_{\mu\sigma,\sigma\rho}
+{1\over 6}\partial_{\mu}F^{a}_{\mu\rho,\sigma\sigma} )
-\eta_{\nu\rho} ({1\over 3}\partial_{\mu}F^{a}_{\mu\sigma,\sigma\lambda}
+{1\over 6}\partial_{\mu}F^{a}_{\mu\lambda,\sigma\sigma} )+\nn\\
+ \eta_{\lambda\rho}({1\over 2} \partial_{\mu}F^{a}_{\mu\nu,\sigma\sigma}
- {1\over 3}\partial_{\mu}F^{a}_{\mu\sigma,\sigma\nu}
+{1\over 3}\partial_{\mu}F^{a}_{\nu\sigma,\sigma\mu})=
j^{a}_{\nu \lambda \rho}.
\eeqa
As we demonstrated, this equation is invariant with respect to the following
gauge transformations:
$$
\delta A^{a}_{\mu\nu\lambda} =\partial_{\mu} \xi^{a}_{\nu\lambda},~~~~~~~~~
\tilde{\delta} A^{a}_{\mu \nu\lambda} =\partial_{\nu} \zeta^{a}_{\mu\lambda}+
\partial_{\lambda} \zeta^{a}_{\mu\nu},
$$
where the gauge parameters are totally symmetric tensors satisfying the
condition (\ref{restrictionongaugeparameters}).
The initial invariance of the equation
$\delta A^{a}_{\mu\nu\lambda} =\partial_{\mu} \xi^{a}_{\nu\lambda}$
imposes restriction on the current $j^{a}_{\nu \lambda \rho}$, in particular, on
its conservation over the first index:
\be\label{tensorcurrentconservation}
\partial_{\nu}j^{a}_{\nu \lambda \rho}=0.
\ee
There are also additional constraints on the current which follow from the
enhanced invariance $\tilde{\delta} A^{a}_{\mu \nu\lambda} =\partial_{\nu} \zeta^{a}_{\mu\lambda}+
\partial_{\lambda} \zeta^{a}_{\mu\nu}$. In Fourier components the
constraints on the group parameters are
\beqa
\omega \zeta_{03} + \kappa \zeta_{33}+
\kappa (\zeta_{00}- \zeta_{11}-\zeta_{22}-\zeta_{33})=0,\nn\\
\omega \zeta_{01} + \kappa \zeta_{31}=0,\nn\\
\omega \zeta_{02} + \kappa  \zeta_{32} =0,\nn\\
\omega \zeta_{00}+ \kappa \zeta_{30} -
\omega (\zeta_{00}- \zeta_{11}-\zeta_{22}-\zeta_{33})=0,\nn
\eeqa
where $k^{\mu}=(\omega,0,0,\kappa)$, therefore
\beqa
\zeta_{00}=
\zeta_{11}+\zeta_{22} -{\omega \over \kappa } \zeta_{03} , \\
\zeta_{31}= - {\omega \over \kappa }\zeta_{01},\nn\\
\zeta_{32} = - {\omega \over \kappa }\zeta_{02},\nn\\
\zeta_{33}= - \zeta_{11}-\zeta_{22} -{\kappa \over \omega }\zeta_{03} ,\nn
\eeqa
and we have six independent gauge parameters
$$
\zeta_{01},\zeta_{02}, \zeta_{03}, \zeta_{11}, \zeta_{22}, \zeta_{12}.
$$
From this it follows that the current components fulfill the following six
relations:
\beqa\label{currentconservation}
k_{\lambda}(j_{1\lambda 0} + j_{0\lambda 1}  +{\omega \over \kappa }
 j_{1\lambda 3} + {\omega \over \kappa } j_{3 \lambda 1})=0, \\
k_{\lambda}(j_{2\lambda 0} + j_{0\lambda 2}  + {\omega \over \kappa }
 j_{2\lambda 3} +  {\omega \over \kappa } j_{3 \lambda 2})=0,\nn\\
 k_{\lambda}( {\omega \over \kappa }
 j_{0\lambda 0} +  {\kappa  \over \omega} j_{3 \lambda 3})=0,\nn\\
k_{\lambda}(j_{0\lambda 0} - j_{3\lambda 3}  + j_{1\lambda 1})=0,\nn\\
k_{\lambda}(j_{0\lambda 0} - j_{3\lambda 3}  + j_{2\lambda 2})=0,\nn\\
k_{\lambda}(j_{1\lambda 2}  + j_{2\lambda 1})=0,\nn
\eeqa
One can also use a different set of independent parameters, in particular:
$
\zeta_{11},\zeta_{12}, \zeta_{13},\zeta_{22}, \zeta_{23}, \zeta_{33}.
$

\section{\it Schwinger Equation for rank-3 Gauge Field}

The Schwinger equation  for symmetric massless rank-3 tensor
gauge field has the form \cite{schwinger}
\beqa\label{schwinger1}
&+&\partial^{2} A_{\alpha \alpha^{'}\alpha^{''}}
-\partial_{\alpha} \partial_{\rho} A_{\rho\alpha^{'}\alpha^{''}}
-\partial_{\alpha^{'}} \partial_{\rho}A_{\alpha\rho\alpha^{''}}
-\partial_{\alpha^{''}} \partial_{\rho}A_{\alpha\alpha^{'}\rho}+\nn\\
&+&\partial_{\alpha}\partial_{\alpha^{'}} A_{\alpha^{''}\rho\rho}
+\partial_{\alpha}\partial_{\alpha^{''}} A_{\alpha^{'}\rho\rho}
+\partial_{\alpha^{'}}\partial_{\alpha^{''}} A_{\alpha\rho\rho}
-3\partial_{\alpha}\partial_{\alpha^{'}}\partial_{\alpha^{''}} A -\nn\\
&-&\eta_{\alpha\alpha^{'}} (\partial^{2}A_{\rho\rho\alpha^{''}}-
\partial_{\lambda}\partial_{\rho} A_{\rho \lambda\alpha^{''}}+{1 \over 2}
\partial_{\alpha^{''}}\partial_{\rho} A_{\rho\lambda \lambda})-\nn\\
&-&\eta_{\alpha\alpha^{''}} (\partial^{2}A_{\rho\rho\alpha^{'}}-
\partial_{\lambda}\partial_{\rho} A_{\rho \lambda\alpha^{'}}+{1 \over 2}
\partial_{\alpha^{'}}\partial_{\rho} A_{\rho\lambda \lambda})-\nn\\
&-&\eta_{\alpha^{'}\alpha^{''}}( \partial^{2} A_{\rho\rho\alpha}
- \partial_{\lambda} \partial_{\rho} A_{\rho\lambda\alpha}
+{1 \over 2}\partial_{\alpha} \partial_{\rho} A_{\lambda\lambda\rho})
=j_{\alpha \alpha^{'}\alpha^{''}}
\eeqa
and contains the scalar field $A$ which should satisfy the high-order differential equation
\beqa
\partial^{2} \partial^{2} A-
\partial^{2}\partial_{\lambda}  A_{\lambda\rho\rho}
+{2\over 3}\partial_{\alpha}\partial_{\lambda}\partial_{\rho}
A_{\alpha \lambda\rho}=0.
\eeqa
Taking derivatives of the l.h.s. of the above equation one can get convinced that
we have conservation of the totally symmetric current $j_{\alpha \alpha^{'}\alpha^{''}}$
\be
\partial_{\alpha} j_{\alpha \alpha^{'}\alpha^{''}}=0
\ee
and the invariance of the equation with respect to the full gauge transformation
\be\label{unrestricted}
\delta A_{\mu \nu\lambda} = \partial_{\mu} \xi_{\nu\lambda}+  \partial_{\nu} \xi_{\mu\lambda}+
\partial_{\lambda} \xi_{\mu\nu}
\ee
without any restrictions on the symmetric gauge parameter $\xi_{\nu\lambda}$.
The great advantage of this formulation is that we have conservation of current
and full gauge symmetry (\ref{unrestricted}).
The disadvantage of this formulation is the appearance of the scalar field $A$
and its high-order differential equation. The illuminating remark of Schwinger
was to make a change of field variable of the form \cite{schwinger}
$$
A_{\alpha \alpha^{'}\alpha^{''}} \rightarrow A_{\alpha \alpha^{'}\alpha^{''}}
-3\partial^{-2}\partial_{\alpha}\partial_{\alpha^{'}}\partial_{\alpha^{''}} A,
$$
which allows to eliminate the scalar field A from the field equation without changing its actual form!
The equation will take a unique form \cite{schwinger}:
\beqa\label{schwinger2}
&+&\partial^{2} A_{\alpha \alpha^{'}\alpha^{''}}
-\partial_{\alpha} \partial_{\rho} A_{\rho\alpha^{'}\alpha^{''}}
-\partial_{\alpha^{'}} \partial_{\rho}A_{\alpha\rho\alpha^{''}}
-\partial_{\alpha^{''}} \partial_{\rho}A_{\alpha\alpha^{'}\rho}+\nn\\
&+&\partial_{\alpha}\partial_{\alpha^{'}} A_{\alpha^{''}\rho\rho}
+\partial_{\alpha}\partial_{\alpha^{''}} A_{\alpha^{'}\rho\rho}
+\partial_{\alpha^{'}}\partial_{\alpha^{''}} A_{\alpha\rho\rho} -\nn\\
&-&\eta_{\alpha\alpha^{'}} (\partial^{2}A_{\rho\rho\alpha^{''}}-
\partial_{\lambda}\partial_{\rho} A_{\rho \lambda\alpha^{''}}+{1 \over 2}
\partial_{\alpha^{''}}\partial_{\rho} A_{\rho\lambda \lambda})-\nn\\
&-&\eta_{\alpha\alpha^{''}} (\partial^{2}A_{\rho\rho\alpha^{'}}-
\partial_{\lambda}\partial_{\rho} A_{\rho \lambda\alpha^{'}}+{1 \over 2}
\partial_{\alpha^{'}}\partial_{\rho} A_{\rho\lambda \lambda})-\nn\\
&-&\eta_{\alpha^{'}\alpha^{''}}( \partial^{2} A_{\rho\rho\alpha}
- \partial_{\lambda} \partial_{\rho} A_{\rho\lambda\alpha}
+{1 \over 2}\partial_{\alpha} \partial_{\rho} A_{\lambda\lambda\rho})
=j_{\alpha \alpha^{'}\alpha^{''}},
\eeqa
but it is not invariant any more with respect to the unrestricted gauge transformations
(\ref{unrestricted}). The gauge parameter should be traceless:
\be\label{tracelesscondition}
\xi_{\mu\mu}=0.
\ee
This leads to the modification of the current conservation law:
\be\label{schwingerconservationlaw}
\partial_{\alpha} j_{\alpha \alpha^{'}\alpha^{''}}
-{1\over 4}\eta_{\alpha^{'}\alpha^{''}}j_{\alpha \rho\rho}=0.
\ee
The conservation law for the current became more sophisticated because of the
traceless restriction on the gauge parameters (\ref{tracelesscondition}).
One can see that the same phenomenon also happens in our case where the
restriction on the gauge parameters $\zeta_{\nu\lambda}$
has the form (\ref{fullgroupofextendedtransformation}) and the
conservation law takes the form (\ref{currentconservation}).
Recent discussion of the Schwinger equation can be found
in \cite{Francia:2005bu,Francia:2002pt}.

I would like to thank  Luis  Alvarez-Gaume, Ignatios  Antoniadis, Ioannis Bakas,
Lars Brink, Ludwig Faddeev, Sergio Ferrara, Peter Minkowski and Raymond Stora for
discussions and CERN Theory Division, where part of this work was completed,
for hospitality.
This work was partially supported by the  EEC Grant no. MRTN-CT-2004-005616.

\section{\it Appendix A}

The invariance of the form $\CL^{'}_3$ can be demonstrated by explicit
variation of each term in the sum (\ref{actionthreeprime}).
Indeed, the variation of the first term is
$$
\delta_{\xi} G^{a}_{\mu\nu,\lambda\rho}G^{a}_{\mu\lambda,\nu\rho}=
2g f^{abc}G^{a}_{\mu\nu,\lambda\rho}G^{b}_{\mu\lambda,\nu}\xi^{c}_{\rho}+
2g f^{abc}G^{a}_{\mu\nu,\lambda\rho}G^{b}_{\mu\lambda,\rho}\xi^{c}_{\nu}+
2g f^{abc}G^{a}_{\mu\nu,\lambda\rho}G^{b}_{\mu\lambda}\xi^{c}_{\nu\rho},
$$
of the second term is
$$
\delta_{\xi} G^{a}_{\mu\nu,\nu\lambda}G^{a}_{\mu\rho,\rho\lambda}=
2g f^{abc}G^{a}_{\mu\nu,\nu\lambda}G^{b}_{\mu\rho,\rho}\xi^{c}_{\lambda}+
2g f^{abc}G^{a}_{\mu\nu,\nu\lambda}G^{b}_{\mu\rho,\lambda}\xi^{c}_{\rho}+
2g f^{abc}G^{a}_{\mu\nu,\nu\lambda}G^{b}_{\mu\rho}\xi^{c}_{\rho\lambda},
$$
of the third term is
\beqa
\delta_{\xi} G^{a}_{\mu\nu,\nu\lambda}G^{a}_{\mu\lambda,\rho\rho}=
2g f^{abc}G^{a}_{\mu\nu,\nu\lambda}G^{b}_{\mu\lambda,\rho}\xi^{c}_{\rho}+
g f^{abc}G^{a}_{\mu\nu,\nu\lambda}G^{b}_{\mu\lambda}\xi^{c}_{\rho\rho}+
g f^{abc}G^{a}_{\mu\lambda,\rho\rho}G^{b}_{\mu\nu,\nu}\xi^{c}_{\lambda}+\nn\\
+g f^{abc}G^{a}_{\mu\lambda,\rho\rho}G^{b}_{\mu\nu,\lambda}\xi^{c}_{\nu}+
g f^{abc}G^{a}_{\mu\lambda,\rho\rho}G^{b}_{\mu\nu}\xi^{c}_{\nu\lambda}\nn,
\eeqa
of the forth term is
\beqa
\delta_{\xi} G^{a}_{\mu\nu,\lambda}G^{a}_{\mu\lambda,\nu\rho\rho}=
g f^{abc}G^{a}_{\mu\lambda,\nu\rho\rho}G^{b}_{\mu\nu}\xi^{c}_{\lambda}+
2g f^{abc}G^{a}_{\mu\lambda,\nu\rho}G^{b}_{\mu\nu,\lambda}\xi^{c}_{\rho}+
g f^{abc}G^{a}_{\mu\lambda,\rho\rho}G^{b}_{\mu\nu,\lambda}\xi^{c}_{\nu}+\nn\\
+g f^{abc}G^{a}_{\mu\nu,\lambda}G^{b}_{\mu\lambda,\nu}\xi^{c}_{\rho\rho}+
2g f^{abc}G^{a}_{\mu\nu,\lambda}G^{b}_{\mu\lambda,\rho}\xi^{c}_{\nu\rho}+
g f^{abc}G^{a}_{\mu\nu,\lambda}G^{b}_{\mu\lambda}\xi^{c}_{\nu\rho\rho}\nn,
\eeqa
of the fifth term is
\beqa
\delta_{\xi} G^{a}_{\mu\nu,\lambda}G^{a}_{\mu\rho,\nu\lambda\rho}=\nn\\
g f^{abc}G^{a}_{\mu\rho,\nu\lambda\rho}G^{b}_{\mu\nu}\xi^{c}_{\lambda}+
g f^{abc}G^{b}_{\mu\rho,\nu\lambda}G^{a}_{\mu\nu,\lambda}\xi^{c}_{\rho}+
g f^{abc}G^{b}_{\mu\rho,\nu\rho}G^{a}_{\mu\nu,\lambda}\xi^{c}_{\lambda}+
g f^{abc}G^{b}_{\mu\rho,\lambda\rho}G^{a}_{\mu\nu,\lambda}\xi^{c}_{\nu}+\nn\\
+g f^{abc}G^{a}_{\mu\nu,\lambda}G^{b}_{\mu\rho,\nu}\xi^{c}_{\lambda\rho}+
g f^{abc}G^{a}_{\mu\nu,\lambda}G^{b}_{\mu\rho,\lambda}\xi^{c}_{\nu\rho}+
g f^{abc}G^{a}_{\mu\nu,\lambda}G^{b}_{\mu\rho,\rho}\xi^{c}_{\nu\lambda}+
g f^{abc}G^{a}_{\mu\nu,\lambda}G^{b}_{\mu\rho}\xi^{c}_{\nu\lambda\rho}\nn,
\eeqa
of the sixth term is
\beqa
\delta_{\xi} G^{a}_{\mu\nu,\nu}G^{a}_{\mu\lambda,\lambda\rho\rho}=\nn\\
g f^{abc}G^{a}_{\mu\lambda,\lambda\rho\rho}G^{b}_{\mu\nu}\xi^{c}_{\nu}+
2g f^{abc}G^{b}_{\mu\lambda,\lambda\rho}G^{a}_{\mu\nu,\nu}\xi^{c}_{\rho}+
g f^{abc}G^{b}_{\mu\lambda,\rho\rho}G^{a}_{\mu\nu,\nu}\xi^{c}_{\lambda}+
g f^{abc}G^{b}_{\mu\lambda,\lambda}G^{a}_{\mu\nu,\nu}\xi^{c}_{\rho\rho}+\nn\\
+2g f^{abc}G^{b}_{\mu\lambda,\rho}G^{a}_{\mu\nu,\nu}\xi^{c}_{\lambda\rho}+
g f^{abc}G^{a}_{\mu\nu,\nu}G^{b}_{\mu\lambda}\xi^{c}_{\lambda\rho\rho}\nn
\eeqa
and finally of the seventh term is
\beqa
\delta_{\xi} G^{a}_{\mu,\nu}G^{a}_{\mu\lambda,\nu\lambda\rho\rho}=\nn\\
2g f^{abc}G^{a}_{\mu\nu}G^{b}_{\mu\lambda,\nu\lambda\rho}\xi^{c}_{\rho}+
g f^{abc}G^{a}_{\mu\nu}G^{b}_{\mu\lambda,\nu\rho\rho}\xi^{c}_{\lambda}+
g f^{abc}G^{a}_{\mu\nu}G^{b}_{\mu\lambda,\lambda\rho\rho}\xi^{c}_{\nu}+
g f^{abc}G^{a}_{\mu\nu}G^{b}_{\mu\lambda,\nu\lambda}\xi^{c}_{\rho\rho}+\nn\\
2g f^{abc}G^{a}_{\mu\nu}G^{b}_{\mu\lambda,\nu\rho}\xi^{c}_{\lambda\rho}+
2g f^{abc}G^{a}_{\mu\nu}G^{b}_{\mu\lambda,\lambda\rho}\xi^{c}_{\nu\rho}+
g f^{abc}G^{a}_{\mu\nu}G^{b}_{\mu\lambda,\rho\rho}\xi^{c}_{\nu\lambda}+\nn\\
g f^{abc}G^{a}_{\mu\nu}G^{b}_{\mu\lambda,\nu}\xi^{c}_{\lambda\rho\rho}+
g f^{abc}G^{a}_{\mu\nu}G^{b}_{\mu\lambda,\lambda}\xi^{c}_{\nu\rho\rho}
+2g f^{abc}G^{a}_{\mu\nu}G^{b}_{\mu\lambda,\rho}\xi^{c}_{\nu\lambda\rho}+
g f^{abc}G^{a}_{\mu\nu}G^{b}_{\mu\lambda}\xi^{c}_{\nu\lambda\rho\rho}.\nn
\eeqa
Some of the terms here are equal to zero, like:
$g f^{abc}G^{a}_{\mu\nu,\lambda}G^{b}_{\mu\rho,\lambda}\xi^{c}_{\nu\rho}$,
$g f^{abc}G^{a}_{\mu\lambda,\lambda}G^{b}_{\mu\nu,\nu}\xi^{c}_{\rho\rho}$
and $g f^{abc}G^{a}_{\mu\nu}G^{b}_{\mu\lambda}\xi^{c}_{\nu\lambda\rho\rho}$.
Amazingly, all nonzero terms cancel each other.

\section{\it Appendix B}

The quadratic form $H^{~'}_{\alpha\alpha^{'}\alpha^{''}\gamma\gamma^{'}\gamma^{''}}$
can be extracted from (\ref{freeactionthreeprime})
and should be symmetrized  over the
$\alpha^{'} \leftrightarrow\alpha^{''}$,
$\gamma^{'} \leftrightarrow \gamma^{''}$ and over the exchange of two sets
of indices
$\alpha \alpha^{'}\alpha^{''} \leftrightarrow \gamma\gamma^{'}\gamma^{''}$,
so that   in the momentum representation it has the form
\beqa
H^{~'}_{\alpha\alpha^{'}\alpha^{''}\gamma\gamma^{'}\gamma^{''}}(k)=
{ k^2 \over 8}\{ &+&\eta_{\alpha\alpha^{'}}
(\eta_{\alpha^{''}\gamma} \eta_{\gamma^{'} \gamma^{''}}
+\eta_{\alpha^{''}\gamma^{'}} \eta_{\gamma \gamma^{''}}
+\eta_{\alpha^{''}\gamma^{''}} \eta_{\gamma\gamma^{'} })\nn\\
&+&\eta_{\alpha\alpha^{''}}
(\eta_{\alpha^{'}\gamma} \eta_{\gamma^{'} \gamma^{''}}
+\eta_{\alpha^{'}\gamma^{'}} \eta_{\gamma \gamma^{''}}
+\eta_{\alpha^{'}\gamma^{''}} \eta_{\gamma\gamma^{'} })\nn\\
&+&\eta_{\alpha\gamma^{'}}
(\eta_{\alpha^{'}\gamma} \eta_{\alpha^{''} \gamma^{''}}
+\eta_{\alpha^{'}\gamma^{''}} \eta_{\alpha^{''}\gamma }
+\eta_{\alpha^{'}\alpha^{''}} \eta_{\gamma\gamma^{''} })\nn\\
&+&\eta_{\alpha\gamma^{''}}
(\eta_{\alpha^{'}\gamma} \eta_{\alpha^{''} \gamma^{'}}
+\eta_{\alpha^{'}\gamma^{'}} \eta_{\alpha^{''}\gamma }+
\eta_{\alpha^{'}\alpha^{''}} \eta_{\gamma \gamma^{'}})~\}\nn\\
             -{ 1\over 8}\{&+&k_{\alpha}k_{\alpha^{'}}
(\eta_{\alpha^{''}\gamma} \eta_{\gamma^{'} \gamma^{''}}
+\eta_{\alpha^{''}\gamma^{'}} \eta_{\gamma\gamma^{''} }
+\eta_{\alpha^{''}\gamma^{''}} \eta_{\gamma\gamma^{'} })\nn\\
&+&k_{\alpha}k_{\alpha^{''}}
(\eta_{\alpha^{'}\gamma} \eta_{\gamma^{'}\gamma^{''} }
+\eta_{\alpha^{'}\gamma^{'}} \eta_{\gamma \gamma^{''}}
+\eta_{\alpha^{'}\gamma^{''}} \eta_{\gamma \gamma^{'}})\nn\\
             &+&k_{\alpha}k_{\gamma^{'}}
(\eta_{\alpha^{'}\gamma} \eta_{\alpha^{''} \gamma^{''}}
+\eta_{\alpha^{'}\gamma^{''}} \eta_{\alpha^{''}\gamma }
+\eta_{\alpha^{'}\alpha^{''}} \eta_{\gamma\gamma^{''} })\nn\\
&+&k_{\alpha}k_{\gamma^{''}}
(\eta_{\alpha^{'}\gamma} \eta_{\alpha^{''} \gamma^{'}}
+\eta_{\alpha^{'}\gamma^{'}} \eta_{\alpha^{''}\gamma }
+\eta_{\alpha^{'}\alpha^{''}} \eta_{\gamma \gamma^{'}})\nn\\
&+&k_{\gamma}k_{\alpha^{'}}
(\eta_{\alpha\gamma^{'}} \eta_{\alpha^{''} \gamma^{''}}
+\eta_{\alpha\gamma^{''}} \eta_{\alpha^{''} \gamma^{'}}
+\eta_{\alpha\alpha^{''}} \eta_{\gamma^{'}\gamma^{''} })\nn\\
&+&k_{\gamma}k_{\alpha^{''}}
(\eta_{\alpha\gamma^{'}} \eta_{\alpha^{'} \gamma^{''}}
+\eta_{\alpha\gamma^{''}} \eta_{\alpha^{'}\gamma^{'} }
+\eta_{\alpha\alpha^{'}} \eta_{\gamma^{'}\gamma^{''} })\nn\\
&+&k_{\gamma}k_{\gamma^{'}}
(\eta_{\alpha\alpha^{'}} \eta_{\alpha^{''} \gamma^{''}}
+\eta_{\alpha\alpha^{''}} \eta_{\alpha^{'} \gamma^{''}}
+\eta_{\alpha\gamma^{''}} \eta_{\alpha^{'}\alpha^{''} })\nn\\
&+&k_{\gamma}k_{\gamma^{''}}
(\eta_{\alpha\alpha^{'}} \eta_{\alpha^{''} \gamma^{'}}
+\eta_{\alpha\alpha^{''}} \eta_{\alpha^{'}\gamma^{'} }
+\eta_{\alpha\gamma^{'}} \eta_{\alpha^{'}\alpha^{''} })~\}\nn\\
   +{ 1\over 4}\{&+&\eta_{\alpha\gamma} (k_{\alpha^{'}}k_{\gamma^{'}}
\eta_{\alpha^{''} \gamma^{''}} + k_{\alpha^{'}}k_{\gamma^{''}}
\eta_{\alpha^{''} \gamma^{'}} + k_{\alpha^{''}}k_{\gamma^{'}}
\eta_{\alpha^{'} \gamma^{''}} \nn\\
&+&k_{\alpha^{''}}k_{\gamma^{''}}
\eta_{\alpha^{'} \gamma^{'}} + k_{\alpha^{'}}k_{\alpha^{''}}
\eta_{ \gamma^{'}\gamma^{''}} +k_{\gamma^{'}}k_{\gamma^{''}}
\eta_{\alpha^{'}\alpha^{''} })~\}.
\eeqa
or combining some of the terms together we shall get an equivalent form
\beqa
H^{~'}_{\alpha\alpha^{'}\alpha^{''}\gamma\gamma^{'}\gamma^{''}}(k)=
{1\over 8}\{ &+& (k^2 \eta_{\alpha\alpha^{'}}-k_{\alpha}k_{\alpha^{'}})
(\eta_{\alpha^{''}\gamma} \eta_{\gamma^{'} \gamma^{''}}
+\eta_{\alpha^{''}\gamma^{'}} \eta_{\gamma \gamma^{''}}
+\eta_{\alpha^{''}\gamma^{''}} \eta_{\gamma\gamma^{'} })\nn\\
&+& (k^2 \eta_{\alpha\alpha^{''}} -k_{\alpha}k_{\alpha^{''}})
(\eta_{\alpha^{'}\gamma} \eta_{\gamma^{'} \gamma^{''}}
+\eta_{\alpha^{'}\gamma^{'}} \eta_{\gamma \gamma^{''}}
+\eta_{\alpha^{'}\gamma^{''}} \eta_{\gamma\gamma^{'} })\nn\\
&+&(k^2 \eta_{\alpha\gamma^{'}}-k_{\alpha}k_{\gamma^{'}})
(\eta_{\alpha^{'}\gamma} \eta_{\alpha^{''} \gamma^{''}}
+\eta_{\alpha^{'}\gamma^{''}} \eta_{\alpha^{''}\gamma }
+\eta_{\alpha^{'}\alpha^{''}} \eta_{\gamma\gamma^{''} })\nn\\
&+&(k^2 \eta_{\alpha\gamma^{''}}-k_{\alpha}k_{\gamma^{''}})
(\eta_{\alpha^{'}\gamma} \eta_{\alpha^{''} \gamma^{'}}
+\eta_{\alpha^{'}\gamma^{'}} \eta_{\alpha^{''}\gamma }+
\eta_{\alpha^{'}\alpha^{''}} \eta_{\gamma \gamma^{'}})~\}\nn\\
             -{ 1\over 8}\{&+&k_{\gamma}k_{\alpha^{'}}
(\eta_{\alpha\gamma^{'}} \eta_{\alpha^{''} \gamma^{''}}
+\eta_{\alpha\gamma^{''}} \eta_{\alpha^{''} \gamma^{'}}
+\eta_{\alpha\alpha^{''}} \eta_{\gamma^{'}\gamma^{''} })\nn\\
&+&k_{\gamma}k_{\alpha^{''}}
(\eta_{\alpha\gamma^{'}} \eta_{\alpha^{'} \gamma^{''}}
+\eta_{\alpha\gamma^{''}} \eta_{\alpha^{'}\gamma^{'} }
+\eta_{\alpha\alpha^{'}} \eta_{\gamma^{'}\gamma^{''} })\nn\\
&+&k_{\gamma}k_{\gamma^{'}}
(\eta_{\alpha\alpha^{'}} \eta_{\alpha^{''} \gamma^{''}}
+\eta_{\alpha\alpha^{''}} \eta_{\alpha^{'} \gamma^{''}}
+\eta_{\alpha\gamma^{''}} \eta_{\alpha^{'}\alpha^{''} })\nn\\
&+&k_{\gamma}k_{\gamma^{''}}
(\eta_{\alpha\alpha^{'}} \eta_{\alpha^{''} \gamma^{'}}
+\eta_{\alpha\alpha^{''}} \eta_{\alpha^{'}\gamma^{'} }
+\eta_{\alpha\gamma^{'}} \eta_{\alpha^{'}\alpha^{''} })~\}\nn\\
   +{ 1\over 4}\{&+&\eta_{\alpha\gamma} (k_{\alpha^{'}}k_{\gamma^{'}}
\eta_{\alpha^{''} \gamma^{''}} + k_{\alpha^{'}}k_{\gamma^{''}}
\eta_{\alpha^{''} \gamma^{'}} + k_{\alpha^{''}}k_{\gamma^{'}}
\eta_{\alpha^{'} \gamma^{''}} \nn\\
&+&k_{\alpha^{''}}k_{\gamma^{''}}
\eta_{\alpha^{'} \gamma^{'}} + k_{\alpha^{'}}k_{\alpha^{''}}
\eta_{ \gamma^{'}\gamma^{''}} +k_{\gamma^{'}}k_{\gamma^{''}}
\eta_{\alpha^{'}\alpha^{''} })~\}.
\eeqa
This expression can be used to calculate divergences. Indeed,
\beqa
k_{\alpha^{'}}H^{~'}_{\alpha\alpha^{'}\alpha^{''}\gamma\gamma^{'}\gamma^{''}}(k)=
{1\over 8}\{&+&(k^2 \eta_{\alpha\alpha^{''}} -k_{\alpha}k_{\alpha^{''}})
(k_{\gamma} \eta_{\gamma^{'} \gamma^{''}}
+k_{\gamma^{'}} \eta_{\gamma \gamma^{''}}
+k_{\gamma^{''}} \eta_{\gamma\gamma^{'} })\nn\\
&+&(k^2 \eta_{\alpha\gamma^{'}}-k_{\alpha}k_{\gamma^{'}})
(k_{\gamma} \eta_{\alpha^{''} \gamma^{''}}
+k_{\gamma^{''}} \eta_{\alpha^{''}\gamma }
+k_{\alpha^{''}} \eta_{\gamma\gamma^{''} })\nn\\
&+&(k^2 \eta_{\alpha\gamma^{''}}-k_{\alpha}k_{\gamma^{''}})
(k_{\gamma} \eta_{\alpha^{''} \gamma^{'}}
+k_{\gamma^{'}} \eta_{\alpha^{''}\gamma }+
k_{\alpha^{''}} \eta_{\gamma \gamma^{'}})~\}\nn\\
             -{ 1\over 8}\{&+&k^{2}k_{\gamma}
(\eta_{\alpha\gamma^{'}} \eta_{\alpha^{''} \gamma^{''}}
+\eta_{\alpha\gamma^{''}} \eta_{\alpha^{''} \gamma^{'}}
+\eta_{\alpha\alpha^{''}} \eta_{\gamma^{'}\gamma^{''} })\nn\\
&+&k_{\gamma}k_{\alpha^{''}}
(2k_{\gamma^{''}}\eta_{\alpha\gamma^{'}}
+2k_{\gamma^{'} }\eta_{\alpha\gamma^{''}}
+k_{\alpha} \eta_{\gamma^{'}\gamma^{''} })\nn\\
&+&k_{\gamma}k_{\gamma^{'}}
(k_{\alpha} \eta_{\alpha^{''} \gamma^{''}}
+2k_{\gamma^{''}}\eta_{\alpha\alpha^{''}} )
+k_{\gamma}k_{\gamma^{''}}k_{\alpha} \eta_{\alpha^{''} \gamma^{'}} ~\}\nn\\
  + { 1\over 4}\{&+&\eta_{\alpha\gamma} (k^{2}k_{\gamma^{'}}
\eta_{\alpha^{''} \gamma^{''}} + k^{2}k_{\gamma^{''}}
\eta_{\alpha^{''} \gamma^{'}} +k^{2} k_{\alpha^{''}}
\eta_{ \gamma^{'}\gamma^{''}} +3k_{\alpha^{''}}k_{\gamma^{'}}k_{\gamma^{''}} ~\}\nn
\eeqa
or using the operator
$H_{\alpha\gamma}= k^2 \eta_{\alpha\gamma} - k_{\alpha}k_{\gamma}$ one can get
\beqa
k_{\alpha^{'}}H^{~'}_{\alpha\alpha^{'}\alpha^{''}\gamma\gamma^{'}\gamma^{''}}(k)=
{1\over 8}\{&+&H_{\alpha\alpha^{''}}~
(k_{\gamma} \eta_{\gamma^{'} \gamma^{''}}
+k_{\gamma^{'}} \eta_{\gamma \gamma^{''}}
+k_{\gamma^{''}} \eta_{\gamma\gamma^{'} })\nn\\
&+&H_{\alpha\gamma^{'}} ~
(k_{\gamma} \eta_{\alpha^{''} \gamma^{''}}
+k_{\gamma^{''}} \eta_{\alpha^{''}\gamma }
+k_{\alpha^{''}} \eta_{\gamma\gamma^{''} })\nn\\
&+&H_{\alpha\gamma^{''}} ~
(k_{\gamma} \eta_{\alpha^{''} \gamma^{'}}
+k_{\gamma^{'}} \eta_{\alpha^{''}\gamma }+
k_{\alpha^{''}} \eta_{\gamma \gamma^{'}})~\}\nn\\
             -{ 1\over 8}\{&+&H_{\alpha\alpha^{''}} ~
              k_{\gamma} \eta_{\gamma^{'}\gamma^{''} }+ H_{\alpha\gamma^{'}} ~
              k_{\gamma} \eta_{\alpha^{''} \gamma^{''}}+H_{\alpha\gamma^{''}}~
              k_{\gamma} \eta_{\alpha^{''} \gamma^{'}}\nn\\
&+&k_{\gamma}k_{\alpha^{''}}
(2k_{\gamma^{''}}\eta_{\alpha\gamma^{'}}
+2k_{\gamma^{'} }\eta_{\alpha\gamma^{''}}
+2k_{\alpha} \eta_{\gamma^{'}\gamma^{''} })\nn\\
&+&k_{\gamma}k_{\gamma^{'}}
(2k_{\alpha} \eta_{\alpha^{''} \gamma^{''}}
+2k_{\gamma^{''}}\eta_{\alpha\alpha^{''}} )
+2k_{\gamma}k_{\gamma^{''}}k_{\alpha} \eta_{\alpha^{''} \gamma^{'}} ~\}\nn\\
  + { 1\over 4}\{&+&\eta_{\alpha\gamma} (k^{2}k_{\gamma^{'}}
\eta_{\alpha^{''} \gamma^{''}} + k^{2}k_{\gamma^{''}}
\eta_{\alpha^{''} \gamma^{'}} +k^{2} k_{\alpha^{''}}
\eta_{ \gamma^{'}\gamma^{''}} +3k_{\alpha^{''}}k_{\gamma^{'}}k_{\gamma^{''}} ~\}\nn
\eeqa
and canceling the identical terms we shall get
\beqa
k_{\alpha^{'}}H^{~'}_{\alpha\alpha^{'}\alpha^{''}\gamma\gamma^{'}\gamma^{''}}(k)=
{1\over 8}\{&+&H_{\alpha\alpha^{''}}
(k_{\gamma^{'}} \eta_{\gamma \gamma^{''}}
+k_{\gamma^{''}} \eta_{\gamma\gamma^{'} })\nn\\
&+&H_{\alpha\gamma^{'}}
(k_{\gamma^{''}} \eta_{\alpha^{''}\gamma }
+k_{\alpha^{''}} \eta_{\gamma\gamma^{''} })\nn\\
&+&H_{\alpha\gamma^{''}}
(k_{\gamma^{'}} \eta_{\alpha^{''}\gamma }+
k_{\alpha^{''}} \eta_{\gamma \gamma^{'}})~\}\nn\\
             -{ 1\over 4}\{
&+&k_{\gamma}k_{\alpha^{''}}
(k_{\gamma^{''}}\eta_{\alpha\gamma^{'}}
+k_{\gamma^{'} }\eta_{\alpha\gamma^{''}})
+k_{\gamma}k_{\gamma^{'}}
k_{\gamma^{''}}\eta_{\alpha\alpha^{''}} ~\}\nn\\
   +{ 1\over 4}\{&+&H_{\alpha\gamma}
   k_{\gamma^{'}}\eta_{\alpha^{''} \gamma^{''}}+H_{\alpha\gamma}
   k_{\gamma^{''}}\eta_{\alpha^{''} \gamma^{'}}+H_{\alpha\gamma}  k_{\alpha^{''}}
\eta_{ \gamma^{'}\gamma^{''}}\nn\\
&+&3\eta_{\alpha\gamma}k_{\alpha^{''}}k_{\gamma^{'}}k_{\gamma^{''}} ~\}.\nn
\eeqa
Again collecting terms we shall get the final expression:
\beqa
k_{\alpha^{'}}H^{~'}_{\alpha\alpha^{'}\alpha^{''}\gamma\gamma^{'}\gamma^{''}}(k)=
{1\over 8}\{&+&H_{\alpha\alpha^{''}}
(k_{\gamma^{'}} \eta_{\gamma \gamma^{''}}
+k_{\gamma^{''}} \eta_{\gamma\gamma^{'} })\nn\\
&+&H_{\alpha\gamma^{'}}
(k_{\gamma^{''}} \eta_{\alpha^{''}\gamma }
+k_{\alpha^{''}} \eta_{\gamma\gamma^{''} })\nn\\
&+&H_{\alpha\gamma^{''}}
(k_{\gamma^{'}} \eta_{\alpha^{''}\gamma }+
k_{\alpha^{''}} \eta_{\gamma \gamma^{'}})~\} \\
             -{ 1\over 4}\{
&+&k_{\gamma}k_{\alpha^{''}}
(k_{\gamma^{''}}\eta_{\alpha\gamma^{'}}
+k_{\gamma^{'} }\eta_{\alpha\gamma^{''}})
+k_{\gamma}k_{\gamma^{'}}
k_{\gamma^{''}}\eta_{\alpha\alpha^{''}}
-3\eta_{\alpha\gamma}k_{\alpha^{''}}k_{\gamma^{'}}k_{\gamma^{''}} ~\}\nn\\
   +{ 1\over 4}\{&+&H_{\alpha\gamma}
   (k_{\gamma^{'}}\eta_{\alpha^{''} \gamma^{''}}+
   k_{\gamma^{''}}\eta_{\alpha^{''} \gamma^{'}}+ k_{\alpha^{''}}
\eta_{ \gamma^{'}\gamma^{''}})~\}\nn,
\eeqa
which has been used in the main text.

\vfill
\end{document}